\documentclass[12pt,a4paper]{article}
\usepackage{a4wide}
\usepackage{latexsym}
\usepackage{epsf}
\usepackage{amssymb}

\makeatletter
\@addtoreset{equation}{section}
\makeatother

\begin{document}

\begin{flushright}
\small
IFT-UAM/CSIC-02-18\\
{\bf hep-th/0210011}\\
October 1st 2002\\
\normalsize
\end{flushright}

\begin{center}


\vspace{.7cm}

{\Large {\bf Gauged/Massive Supergravities in\\ Diverse Dimensions}}\\

\vspace{1.2cm}

{\bf\large Natxo Alonso-Alberca}${}^{\spadesuit,\heartsuit,}$
\footnote{E-mail: {\tt natxo@leonidas.imaff.csic.es}}
{\large and}
{\bf\large Tom\'as Ort\'{\i}n}${}^{\spadesuit,\clubsuit,}$
\footnote{E-mail:  {\tt Tomas.Ortin@cern.ch} }
\vskip 1truecm

${}^{\spadesuit}$\ {\it Instituto de F\'{\i}sica Te\'orica, C-XVI,
Universidad Aut\'onoma de Madrid \\
E-28049-Madrid, Spain}

\vskip 0.2cm
${}^{\heartsuit}$\ {\it Departamento de F\'{\i}sica Te{\'o}rica, C-XI,
Universidad Aut\'onoma de Madrid\\
Cantoblanco, E-28049 Madrid, Spain}

\vskip 0.2cm
${}^{\clubsuit}$\ {\it I.M.A.F.F., C.S.I.C., Calle de Serrano 113 bis\\ 
E-28006-Madrid, Spain}

\vspace{.7cm}


{\bf Abstract}

\end{center}

\begin{quotation}

\small

We show how massive/gauged maximal supergravities in $11-n$ dimensions
with $SO(n-l,l)$ gauge groups (and other non-semisimple subgroups of
$Sl(n,\mathbb{R})$) can be systematically obtained by dimensional
reduction of ``massive 11-dimensional supergravity''. This series of
massive/gauged supergravities includes, for instance, Romans' massive
$N=2A,d=10$ supergravity for $n=1$, $N=2,d=9$ $SO(2)$ and $SO(1,1)$
gauged supergravities for $n=2$, and $N=8,d=5$ $SO(6-l,l)$ gauged
supergravity.  In all cases, higher $p$-form fields get masses through
the St\"uckelberg mechanism which is an alternative to self-duality in
odd dimensions.

\end{quotation}

\newpage

\pagestyle{plain}

\section{Introduction}

Massive/gauged supergravities are very interesting theories which have
become fashionable due to the relation between the presence in their
Lagrangians of mass/gauge coupling parameters with the existence of
domain-wall-type solutions and the holographic relation between (in
general non-conformal) the gauge field theories that live on the
domain wall and the superstring/supergravity theories that live in the
bulk \cite{Maldacena:1997re,Boonstra:1998mp} (for a review, see
Ref.~\cite{Aharony:1999ti}).

These supergravity theories appear in the literature in essentially 3
different ways in the compactification of ungauged (massless)
supergravities:

\begin{enumerate}
  
\item In compactifications in non-trivial internal manifolds
  (particularly Freund-Rubin-type \cite{Freund:1980xh} spontaneous
  compactifications on spheres).  Some notable examples are the
  $S^{7}$ \cite{Duff:1983gq} compactification of 11-dimensional
  supergravity that is supposed to give the $SO(8)$-gauged $N=8,d=4$
  supergravity, the $S^{4}$ compactification of 11-dimensional
  supergravity \cite{Pilch:xy} that gives \cite{Nastase:1999cb} the
  $SO(5)$-gauged $N=4,d=7$ theory, and the $S^{5}$ compactification of
  $N=2B,d=10$ supergravity \cite{Schwarz:qr,Gunaydin:1984fk,Kim:1985ez}
  that gives the $SO(6)$-gauged $N=8,d=5$ supergravity theory.
  
\item In Scherk-Schwarz generalized dimensional reductions
  \cite{Scherk:1979zr}, in which the global symmetry is geometrical or
  non-geometrical.
  
  Examples in which a global geometrical $SU(2)$ symmetry has been
  used to obtain a gauged/massive supergravity are Salam \& Sezgin's
  compactification of 11-dimensional supergravity to obtain
  $SU(2)$-gauged $N=4,d=8$ supergravity \cite{Salam:1984ft}, and
  Chamseddine and Volkov's obtention of $SU(2)\times SU(2)$-gauged
  $N=4,d=4$ supergravity \cite{Chamseddine:1997nm,Chamseddine:1997mc}
  and Chamseddine and Sabra's obtention of and $SU(2)$-gauged
  $N=2,d=7$ supergravity \cite{Chamseddine:1999uy} from $N=1,d=10$
  supergravity in both cases.
  
  An example in which a global symmetry of non-geometrical origin is
  used to obtain a gauged/massive supergravity by generalized
  dimensional reduction is the obtention of massive $N=2,d=9$
  supergravity from $N=2B,d=10$ supergravity exploiting the axion's
  shift symmetry \cite{Bergshoeff:1996ui}. If one exploits the full
  global $Sl(2,\mathbb{R})$ symmetry of the $N=2B,d=10$ theory one
  obtains a 3-parameter family of supergravity theories
  \cite{Meessen:1998qm} (see also \cite{Gheerardyn:2001jj}) some of
  which are gauged supergravities \cite{Cowdall:2000sq}.  From the
  string theory point of view, the three parameters take discrete
  values which must be considered equivalent when they are related by
  an $Sl(2,\mathbb{Z})$ duality transformation (i.e.~when they belong
  to the same conjugacy class) and they describe the low-energy limit
  of the same string theory \cite{Hull:2002wg}. There is, actually, an
  infinite number of $Sl(2,\mathbb{Z})$ conjugacy classes and for each
  of them one gets a massive/gauged supergravity with either $SO(2)$,
  $SO(1,1)$ or no gauge group \cite{Bergshoeff:2002mb}\footnote{These
    gauged supergravity theories have been constructed by direct
    gauging in Ref.~\cite{Nishino:2002zi}. Further, very recently, new
    gauged $N=2,d=9$ supergravities constructed by Scherck-Schwarz
    generalized dimensional reduction have been presented in
    Ref.~\cite{Bergshoeff:2002nv}. One of them has a 2-dimensional
    non-Abelian gauge group.}. We will take a closer look later to
  this $Sl(2,\mathbb{Z})$ family of theories.

\item In compactifications with non-trivial $p$-form fluxes (see
  e.g.~\cite{Louis:2002ny}).

\end{enumerate}

These three instances are not totally unrelated.  To start with,
compactifications with fluxes can be understood as non-geometrical
Scherk-Schwarz reductions in which the global symmetry exploited is
the one generated by $p$-form ``gauge'' transformations with constant
parameters. The axion shift symmetry can be understood as the limit
case $p=-1$ and can be used in compactifications on circles. Higher
$p$-form fluxes can only be exploited in higher-dimensional internal
spaces that can support them.  On the other hand, the geometrical
Scherk-Schwarz compactifications used by Salam \& Sezgin and
Chamseddine, Volkov and Sabra could be understood as compactification
on the $SU(2)$ group manifold $S^{3}$ although one would expect a
gauge group $SO(4)\sim SU(2) \times SU(2)$ since this is the isometry
group of the $S^{3}$ metric used.

Finally, the Freund-Rubin spontaneous sphere compactifications are
compactifications on a brane background (more precisely, in a brane's
near-horizon geometry) and there is a net flux of the form associated
to the brane, while the non-geometrical Scherk-Schwarz
compactifications can also be seen as compactifications on a
$(d-3)$-brane background \cite{Meessen:1998qm}, in which the brane
couples to the $(d-2)$-form potential dual to the scalar.

Historically, almost all the gauged/massive theories we just discussed
had been constructed by gauging or mass-deforming known
ungauged/massless theories\footnote{The $SO(8)$-gauged $N=8,d=4$
  supergravity was constructed in
  Ref.~\cite{deWit:1981eq,deWit:1982ig}, the $SO(5)$-gauged $N=4,d=7$
  supergravity in Ref.~\cite{Pernici:xx}, the $SO(6)$-gauged $N=8,d=5$
  supergravity in Refs.~\cite{Gunaydin:1984qu,Pernici:ju}, the
  $SU(2)\times SU(2)$-gauged $N=4,d=4$ supergravity in
  Ref.~\cite{Freedman:ra} and the $SU(2)$-gauged $N=2,d=7$
  supergravity in Ref.~\cite{Townsend:1983kk}. Romans' massive
  $N=2A,d=10$ supergravity was obtained by a deformation of the
  massless theory \cite{Romans:tz}.}  the only exception being the
$N=2,d=9$ theories that, in principle, could have been constructed in
that way as well.

A crucial ingredient in the gauging of some of the higher-dimensional
supergravities with $p$-form fields transforming under the global
symmetry being gauged is that these fields must be given a mass whose
value is related by supersymmetry to the gauge coupling parameter: if
the $p$-form fields remained massless, they should transform
simultaneously under their own massless $p$-form gauge transformations
(to decouple negative-norm states) and under the new gauge
transformations, which is impossible. A mass term eliminates the
requirement of massless $p$-form gauge invariance but introduces
another problem, because the number of degrees of freedom of the
theory should remain invariant. In the cases of the $SO(5)$-gauged
$N=4,d=7$ and the $SO(6)$-gauged $N=8,d=5$ theories this was achieved
by using the ``self-duality in odd dimensions'' mechanism
\cite{Townsend:xs,Townsend:ad} which we will explain later on.

The need to introduce mass parameters together with the gauge coupling
constant is one of the reasons why we call these theories
gauged/massive supergravities. In some cases no mass parameters will
be needed in the gauging and in some others no gauge symmetry will be
present when the mass paramaters are present but they can nevertheless
be seen as members of the same class of theories. Another reason is
that in many cases the gauge parameter has simultaneously the
interpretation of gauge coupling constant and mass of a domain-wall
solution of the theory\footnote{Anti-de Sitter spacetime can also be
  interpreted as a domain-wall solution, its mass being related to the
  cosmological constant.} that can correspond to the near-horizon
limit of some higher-dimensional brane solution apart from that of the
mass of a given field in the Lagrangian.

The gauging and mass-deformation procedures are very effective tools
to produce gauged theories in a convenient form but hide completely
their possible higher-dimensional or string/M-theorical origin. In
fact, there are many gauged/massive supergravity theories whose
string- or M-Theoretical origin is still unknown, which, in
supergravity language means that we do not know how to obtain them by
some compactification procedure from some higher-dimensional
(ungauged/massless) theory. In some cases, it is known how to obtain
it from $N=2B,d=10$ supergravity but not from the $N=2A,d=10$ or
11-dimensional supergravity. A notorious example is Romans' massive
$N=2A,d=10$ supergravity \cite{Romans:tz} that cannot be obtained from
standard 11-dimensional supergravity (a theory that cannot be deformed
to accomodate a mass parameter preserving 11-dimensional Lorentz
invariance \cite{Bautier:1997yp,Deser:1997gm,Deser:1998xi}) by any
sort of generalized dimensional reduction, but there are many more.
Let us review some other examples:

\begin{enumerate}
  
\item The $Sl(2,\mathbb{Z})$ family of $N=2,d=9$ gauged/massive
  supergravities are obtained by Scherk-Schwarz reduction of the
  $N=2B,d=10$ theory, but it is not known how to obtain them from
  standard 11-dimensional or $N=2A,d=10$ supergravity.
  
  In these theories, the $Sl(2,\mathbb{R})$ doublet of 2-form
  potentials gets masses through the St\"uckelberg mechanism.
  
\item The massless $N=4,d=8$ supergravity contains two $SU(2)$
  triplets of vector fields. The two triplets are related by
  $Sl(2,\mathbb{R})$ S-duality transformations. It should be possible
  to gauge $SU(2)$ using as $SU(2)$ gauge fields any of the two
  triplets. If we gauged the triplet of Kaluza-Klein vectors, we would
  get the theory that Salam \& Sezgin obtained by Scherk-Schwarz
  reduction of 11-dimensional supergravity. It is not known how to
  derive from standard 11-dimensional supergravity the ``S-dual''
  theory that one would get gauging the other triplet, that comes from
  the 11-dimensional 3-form.
  
  In these two theories, the $SU(2)$ triplet of 2-form potentials gets
  masses through the St\"uckelberg mechanism, eating the 3 vectors
  that are not $SU(2)$-gauged.

\item The $SO(5)$-gauged $N=4,d=7$ supergravity theory is also just a
  particular member of the family of $SO(5-l,l)$ gauged $N=4,d=7$
  supergravities constructed in Ref.~\cite{Pernici:zw}. The
  11-dimensional origin of the $SO(5)$ theory is well understood, but
  not that of the theories with non-compact gauge group.
  
\item The $SO(6)$-gauged $N=8,d=5$ supergravity theory is also just a
  particular member of the family of $SO(6-l,l)$ gauged $N=8,d=5$
  supergravities constructed in
  Refs.~\cite{Gunaydin:1984qu,Pernici:ju}. Again, while the
  $N=2B,d=10$ origin of the $SO(6)$ theory is well understood, its
  11-dimensional origin and the higher-dimensional origin of the
  theories with non-compact gauge groups is unknown.
  
\item Essentially the same can be said about the $SO(8)$-gauged
  $N=8,d=4$ supergravity theory since it is possible to generate from
  it by analytical continuation theories with non-compact groups
  $SO(8-l,l)$ \cite{Hull:2002cv} whose higher-dimensional origin is
  also unknown.

\end{enumerate}

In the search for an 11-dimensional origin of Romans' massive
$N=2A,d=10$ theory a ``massive 11-dimensional supergravity'' was
proposed in Ref.~\cite{Bergshoeff:1997ak}. This theory is a
deformation of the standard 11-dimensional supergravity that contains
a mass parameter and, to evade the no-go theorem of
Refs.~\cite{Bautier:1997yp,Deser:1997gm,Deser:1998xi}, a Killing
vector in the Lagrangian that effectively breaks the 11-dimensional
Lorentz symmetry to the 10-dimensional one even if the theory is
formally 11-dimensional covariant.  Standard dimensional reduction in
the direction of the Killing vector gives the Lagrangian of Romans'
theory.

This theory was little more than the straightforward uplift of Romans'
but it could be generalized to one with $n$ Killing vectors and a
symmetric\footnote{We will discuss later on the possibility of
  generalizing further the theory by admitting non-symmetric matrices
  $\mathsf{Q}^{mn}$.} $n\times n$ mass matrix $\mathsf{Q}^{mn}$
\cite{Meessen:1998qm}.  The reduction of the $n=2$ theory in the
direction of the two Killing vectors turns out to give all the
$SO(2-l,l)$-gauged $N=2,d=9$ supergravities obtained by Scherk-Schwarz
reduction from $N=2B,d=10$ supergravity
\cite{Meessen:1998qm,Gheerardyn:2001jj}: each of these theories is
determined by a traceless $2\times 2$ matrix $m^{m}{}_{n}$ of the
$sl(2,\mathbb{R})$ Lie algebra which is related to the symmetric mass
matrix $\mathsf{Q}^{mn}$ by

\begin{displaymath}
\mathsf{Q}^{mn} = \eta^{mp} \mathsf{m}_{p}{}^{n}\, ,
\hspace{1cm}
\eta_{mn} \equiv
\left(
\begin{array}{cc}
 0 & 1 \\
 -1 & 0 \\
\end{array}
\right)\, ,
\hspace{1cm}
\eta^{mn}=-\eta_{mn}\, .
\end{displaymath}

The reduction of the $n=3$ theory gives the ``S-dual'' $SU(2)$-gauged
$N=4,d=8$ theory mentioned above when we make the choice
$\mathsf{Q}^{mn}=g \delta^{mn}$ \cite{Alonso-Alberca:2000gh} but can
also give the theories with non-compact gauge group $SO(2,1)$ if we
choose $\mathsf{Q}=g\, {\rm diag}\,(++-)$.  Singular $\mathsf{Q}$s
give rise other 3-dimensional non-semisimple gauge groups and massive/ungauged
supergravities, as we are going to show.

In this paper we are going to study these and other gauged/massive
theories obtained by dimensional reduction of ``massive 11-dimensional
supergravity'' with $n$ Killing vectors
\cite{Bergshoeff:1997ak,Meessen:1998qm}. Generically, the theories
obtained in this way are $(11-n)$-dimensional supergravity theories
with 32 supercharges determined by a mass matrix $\mathsf{Q}^{mn}$.
They are {\it covariant} under global $Sl(n,\mathbb{R})$ duality
transformations that in general transform $\mathsf{Q}^{mn}$ into the
mass matrix of another theory\footnote{In all cases we expect the
  entries of the mass matrix $\mathsf{Q}^{mn}$ to be quantized and
  take integer values in appropriate units, since they are related to
  tensions and charges of branes which are quantized in string theory.
  The duality group is then broken to $Sl(n,\mathbb{Z})$
  \cite{Hull:1994ys}. Theories related by $Sl(n,\mathbb{Z})$
  transformations should be considered equivalent from the string
  theory point of view. We will take into account these subtleties
  later.} of the same family. 

The subgroup of $Sl(n,\mathbb{R})$ that preserves the mass matrix is a
symmetry of the theory and at the end it will be the gauge group. If
we use $Sl(n,\mathbb{R})$ transformations and rescalings to
diagonalize the mass matrix so it has only $+1,-1,0$ in the diagonal,
it is clear that $SO(n,n-l)$ will be amongst the possible gauge groups
and corresponds to a non-singular mass matrix.  These theories with
non-singular mass matrices have $n(n-1)/2$ vector fields coming from
the $\hat{C}_{\mu mn}$ components of the 11-dimensional 3-form and
transforming as $SO(n-l,l)$ $l=0,\ldots,n$ gauge vector fields plus
$n$ 2-forms with the same origin and $n$ Kaluza-Klein vectors coming
from the 11-dimensional metric that transform as $SO(n-l,l)$ $n$-plets
. The $n$ vectors act as St\"uckelberg fields for the 2-forms which
become massive. In this way the theory is consistent with the
$SO(n-l,l)$ gauge symmetry.

Finally, all these theories have a scalar potential that contains a
universal term of the form

\begin{equation}
\label{eq:universalterm}
 {\cal V} = -{\textstyle\frac{1}{2}} e^{\alpha \varphi}\,
\left\{ \left[ {\rm Tr} (\mathsf{Q}{\cal M})\right]^2 
-2 {\rm Tr}(\mathsf{Q}{\cal M})^{2} \right\}\, ,
\end{equation}

\noindent
where $\mathcal{M}$ is a (symmetric) $Sl(n,\mathbb{R})/SO(n)$ scalar
matrix, plus, possibly, other terms form the scalars that come from
the 3-form. That scalar potentials of this form appears in several
gauged supergravities was already noticed in
Refs.~\cite{Salam:1984ft,Gunaydin:1986cu}. The $d=5$ case is special
because $\alpha=0$. This is related to the invariance of the
Lagrangian under the $N=2B,d=10$ $Sl(2,\mathbb{R})$ symmetry.

Some of these theories are known, albeit in a very different form.
The case $n=6$ is particularly interesting: we get $SO(6-l,l)$-gauged
$N=8,d=5$ supergravities which were constructed by explicit gauging in
Refs.~\cite{Gunaydin:1984qu,Pernici:ju}, with 15 gauge vectors that
originate in the 3-form, 6 Kaluza-Klein vector fields that originate
in the metric and give mass by the St\"uckelberg mechanism to 6
2-forms that come from the 3-form. That is: the field content (but not
the couplings nor the spectrum) is the same as that of the ungauged
theory that one would obtain by straightforward toroidal dimensional
reduction. In fact, the ungauged theory can be recovered by taking the
limit $\mathsf{Q}\rightarrow 0$ which is non-singular. In
Refs.~\cite{Gunaydin:1984qu,Pernici:ju} the gauged theories were
constructed by dualizing first the 6 vectors into 2-forms that,
together with the other 6 2-forms, satisfy self-duality equations
\cite{Townsend:xs} and describe also the degrees of freedom of 6
massive 2-forms. In this theory the massles limit is singular and can
only be taken after the elmination of the 6 unphsyical 2-forms
\cite{Townsend:ad}.

Thus, we have, presumably, two different versions of the same theory
in which the 6 massive 2-forms are described using the St\"uckelberg
formalism or the self-duality formalism. We will try to show the full
equivalence between both formulations at the classical level.

Something similar happens in $d=7$, although we get $SO(4-l,l)$-gauged
theories and in the literature only $SO(5-l,l)$-gauged theories have
been constructed \cite{Pernici:xx,Pernici:zw}.

This paper is organized as follows: in Section~\ref{sec-BLOM} we
describe the ``massive 11-dimensional supergravity'', its Lagrangian
and symmetries. In Section~\ref{sec-Romans} we briefly review how for
$n=1$ we recover Romans' massive $N=2A,d=10$ supergravity. In
Section~\ref{sec-d=9} we revise how for $n=2$ we get the
$Sl(2,\mathbb{Z})$ family of gauged/massive $N=2,d=9$ supergravities
and how they are classified by their mass matrix. In
Section~\ref{sec-d=8} we study the case $n=3$ and the gauged/massive
$N=2,d=8$ supergravities that arise which allows us to describe the
general situation for arbitrary $n$.  In Section~\ref{sec-d=5} we
study the $n=6$ case and try to argue that we have obtained an
alternative but fully equivalent form of the $SO(6-l,l)$-gauged
$N=8,d=5$.

\section{Massive 11-Dimensional Supergravity}
\label{sec-BLOM}

Massive 11-dimensional supergravity can be understood as a deformation of
standard 11-dimensional supergravity \cite{Cremmer:1978km} that breaks
11-dimensional Lorentz invariance. The bosonic fields of the standard
$N=1,d=11$ supergravity are the Elfbein and a 3-form potential\footnote{Our
  conventions are those of Refs.~\cite{Meessen:1998qm,Alonso-Alberca:2000gh}.
  In particular, $\hat{\mu}$ ($\hat{a}$) are curved (flat) 11-dimensional
  indices and our signature is $(+-\cdots-)$.}

\begin{equation}
\left\{\hat{e}_{\hat{\mu}}{}^{\hat{a}},
\hat{C}_{\hat{\mu}\hat{\nu}\hat{\rho}}
\right\}\, .
\end{equation}

\noindent
 The field strength of the 3-form is

\begin{equation}
\hat{G} = 4\, \partial\, \hat{C}\, ,
\end{equation}

\noindent
 and is obviously invariant under the massless 3-form 
gauge transformations

\begin{equation}
\label{eq:3formgauge}
\delta\, \hat{C}= 3\, \partial\, \hat{\chi}\, ,
\end{equation}

\noindent
 where $\hat{\chi}$ is a 2-form. 
The action for these bosonic fields is

\begin{equation}
\label{eq:action11}
 \hat{S} = 
\int d^{11} \hat{x} \sqrt{|\hat{g}|}
 \left[ \hat{R} -{\textstyle\frac{1}{2 \cdot 4!}}
\hat{G}{}^{2}
 -{\textstyle\frac{1}{(144)^{2}}}
{\textstyle\frac{1}{\sqrt{|\hat{g}|}}}\, \hat{\epsilon}
\hat{G} \hat{G}\hat{C} \right]\, .
\end{equation}

Now, we are going to assume that all the fields of this theory are
independent of $n$ internal coordinates $z^{m}$. This is the standard
assumption in toroidal dimensional reductions and, in particular, it
means that the metric admits $n$ mutually commuting Killing vectors
$\hat{k}_{(n)}$ associated to the internal coordinates by 

\begin{equation}
\label{eq:KillDef}
\hat{k}_{(m)}{}^{\hat{\mu}}
\partial_{\hat{\mu}}= \frac{\partial\,\,\,}{\partial z^{m}}
\equiv \partial_{m}\, .
\end{equation}

\noindent
 We are also going to introduce  an arbitrary 
symmetric {\it mass} matrix $\mathsf{Q}^{mn}$. The possibility or need
to introduce a more general mass matrix will be discussed later on.
With these elements (the Killing vectors and the mass matrix) we are
going to deform the massless theory.

First, we construct the {\it massive gauge parameter}
1-form\footnote{$i_{\hat{v}}\hat{T}$ denotes the contraction of the
  last index of the covariant tensor $\hat{T}$ with the vector
  $\hat{v}$.}  $\hat{\lambda}{}^{(n)}$

\begin{equation}
\hat{\lambda}{}^{(m)} 
\equiv -\mathsf{Q}^{mn}i_{\hat{k}_{\scriptscriptstyle (n)}}\!\hat{\chi}\, ,
\end{equation}

\noindent
 and, for any 11-dimensional tensor $\hat{T}$, we
define the {\it massive gauge transformations}

\begin{equation}
\delta_{\hat{\chi}}\hat{T}_{\hat{\mu}_{1}\cdots\hat{\mu}_{r}}\;=\;
-\hat{\lambda}{}^{(n)}{}_{\hat{\mu}_{1}} 
\hat{k}_{(n)}{}^{\hat{\nu}}
\hat{T}_{\hat{\nu}\hat{\mu}_{2}\cdots\hat{\mu}_{r}} 
-\cdots-
\hat{\lambda}{}^{(n)}{}_{\hat{\mu}_{r}} 
\hat{k}_{(n)}{}^{\hat{\nu}}
\hat{T}_{\hat{\mu}_{1}\cdots\hat{\mu}_{\scriptscriptstyle r-1}\hat{\nu}}\, .
\end{equation}

According to this general rule, the massive gauge transformation of
the 11-dimensional metric $\hat{g}_{\hat{\mu}\hat{\nu}}$ and of any
11-dimensional form of rank $r$
$\hat{A}_{\hat{\mu}_{1}\cdots\hat{\mu}_{r}}$ are given by

\begin{equation}
\left\{\begin{array}{rcl}
\delta_{\hat{\chi}}
\hat{g}_{\hat{\mu}\hat{\nu}} & = &
-2 \hat{k}_{(m)\, (\hat{\mu}}\,
\hat{\lambda}{}^{(m)}{}_{\hat{\nu})}\, ,\\
& & \\
\delta_{\hat{\chi}}
\hat{A}_{\hat{\mu}_{1}\cdots\hat{\mu}_{r}} & = &
(-)^{r}r\hat{\lambda}{}^{(n)}{}_{[\hat{\mu}_{1}}\,
\left(i_{\hat{k}_{(n)}}
\hat{A}\right){}_{\hat{\mu}_{2}\cdots\hat{\mu}_{r}]}\, ,\\
\end{array}
\right.
\end{equation}

\noindent
 which, together,  imply

\begin{equation}
\left\{
\begin{array}{lcl}
\delta_{\hat{\chi}} \sqrt{|\hat{g}|} 
& = & 0 \, , \\
& &        \\
\delta_{\hat{\chi}} \hat{A}{}^{2}     
& = & 0 \, .
\end{array}
\right.
\end{equation}

However, the 3-form of massive 11-dimensional supergravity does not
transform homogeneously under massive gauge transformations, but 

\begin{equation}
\delta_{\hat{\chi}} \hat{C} =
3\partial\hat{\chi} 
-3\hat{\lambda}{}^{(n)}
i_{\hat{k}_{\scriptscriptstyle (n)}}\!\hat{C}\, ,
\end{equation}

\noindent
  which allows us to see it as a sort of {\it connection}. 
It is not surprising that the gauge vectors of the dimensionally
reduced gauged/massive theories come from the 11-dimensional 3-form.
The massive 4-form field strength is given by

\begin{equation}
\hat{G}= 
4\partial \hat{C} 
-3\mathsf{Q}^{mn}\,i_{\hat{k}_{\scriptscriptstyle (m)}}\!\hat{C}\,
i_{\hat{k}_{\scriptscriptstyle (n)}}\!\hat{C}\, ,
\end{equation}

\noindent
 and transforms covariantly, according to the above general rule,
so 

\begin{equation}
\delta_{\hat{\chi}}\hat{G}^{2}=0 \, . 
\end{equation}

The action of the proposed {\it massive 11-dimensional supergravity}
then reads

\begin{equation}
\begin{array}{rcl}
\hat{S}
& = &
{\displaystyle\int} d^{11}\hat{x}\sqrt{|\hat{g}|}\,
\left\{
\hat{R}(\hat{g}) 
-\textstyle{\frac{1}{2\cdot 4!}} \hat{G}^{2} 
-\hat{K}_{\hat{\mu}\hat{\nu}\hat{\rho}}
\hat{K}^{\hat{\nu}\hat{\rho}\hat{\mu}}
+\textstyle{\frac{1}{2}}\mathsf{Q}^{mn}
d\hat{k}_{(m)}            
i_{\hat{k}_{(n)}}\!\hat{C} 
\right.\\
& & \\
& &
\hspace{2cm}
+\textstyle{\frac{1}{2}}(\mathsf{Q}^{mn}
\hat{k}_{(m)}{}^{\hat{\mu}}\hat{k}_{(n)\, \hat{\mu}})^{2}
-( \mathsf{Q}^{mn}  \hat{k}_{(m)\, \hat{\mu}}\hat{k}_{(m)\, \hat{\nu}})^{2}
\\
& & \\
& &
\hspace{2cm}
-\frac{1}{6^{4}}
\frac{\hat{\epsilon}}{\sqrt{|\hat{g}|}}
\left[ 
\partial\hat{C}\partial\hat{C}\hat{C}
+ \frac{9}{8}
\mathsf{Q}^{mn}
\partial\hat{C}\hat{C}\, i_{\hat{k}_{(m)}}\!\hat{C}\, 
i_{\hat{k}_{(n)}}\hat{C}
\right. \\
& & \\
& &
\hspace{3.5cm}
\left.
\left.
+ \frac{27}{80}\mathsf{Q}^{mn}\mathsf{Q}^{pq}
\hat{C}\, 
i_{\hat{k}_{(m)}}\!\hat{C}\, 
i_{\hat{k}_{(n)}}\hat{C}\, 
i_{\hat{k}_{(p)}}\!\hat{C}\, 
i_{\hat{k}_{(q)}}\hat{C}
\right]
\right\}\; ,
\end{array}
\label{eq:masact.11}
\end{equation}

\noindent
 where we have defined a {\it contorsion} tensor

\begin{equation}
\hat{K}_{\hat{a}\hat{b}\hat{c}}
= {\textstyle\frac{1}{2}}
\left(
\hat{T}_{\hat{a}\hat{c}\hat{b}}
+\hat{T}_{\hat{b}\hat{c}\hat{a}}
-\hat{T}_{\hat{a}\hat{b}\hat{c}}
\right) \; ,
\label{eq:torsion.2}
\end{equation}

\noindent
 where the {\it torsion} tensor is defined by

\begin{equation}
\hat{T}_{\hat{\mu}\hat{\nu}}{}^{\hat{\rho}}
= -\mathsf{Q}^{mn}(i_{\hat{k}_{(m)}}\!\hat{C})_{\hat{\mu}\hat{\nu}} 
\hat{k}_{(n)}{}^{\hat{\rho}}\, .
\end{equation}

The action is also invariant under massive gauge transformations up to
total derivatives.

By construction, this theory is meant to be compactified in the
$n$-dimensional torus parametrized by the coordinates $z^{m}$. After
that dimensional reduction, the explicit Killing vectors in the
Lagrangian disappear and one gets a genuine $(11-n)$-dimensional field
theory. We will postpone to the last section the discussion of the
``true'' dimensional nature of the above theory and in the next few
sections we will consider it as a systematic prescription to get
massive/gauged supergravity theories in $11-n$ dimensions and we will
study these theories for several values of $n$.

Also by construction, there is a natural action of the group
$Gl(n,\mathbb{R})$ in these theories, all the objects carrying $m,n$
indices (including the mass matrix) transforming in the vector
representation. The subgroup of $Gl(n,\mathbb{R})$ that preserves the
mass matrix will be a symmetry group of the theory.

The gauge invariances of the gauged supergravities that we will obtain
are encoded in the 11-dimensional {\it massive gauge transformations}
parametrized by the 1-forms $\hat{\lambda}^{(m)}$. Their dimensional
reduction will give rise to further massive gauge transformations
parametrized by 1-forms and associated to massive 2-forms
$\lambda^{(m)}_{\mu}$ and will also give rise to (Yang-Mills) gauge
transformations parametrized by the scalars $\lambda^{(m)}_{n}$ where
the subindex $n$ corresponds to an internal direction. These scalars
exist when there is more than one Killing vector and are antisymmetric
in the indices $m,n$ and correspond to orthogonal gauge groups. This
is consistent with the fact that the gauge vectors come from the
components $\hat{C}_{\mu mn}$ and naturally carry a pair of
antisymmetric indices corresponding to the adjoint representation of
an orthogonal group.

A few words about the fermions of the theory, which we have so far
ignored, are in order. The above theory is a straightforward
generalization to arbitrary $n$ of the $n=2$ case obtained by
uplifting of the gauged/massive $N=2,d=9$ supergravities constructed
in Ref.~\cite{Meessen:1998qm} by non-geometrical Scherk-Schwarz
reduction of the $N=2B,d=10$ theory. This construction was made for
the bosonic sector only, but can be made for the full supergravity
Lagrangian, as shown in Ref.~\cite{Gheerardyn:2001jj}. Once the full
gauged/massive $N=2,d=9$ supergravity is constructed it can be
uplifted to $d=11$ and then generalized to arbitrary $n$. This was
done for the fermionic supersymmetry transformation rules in
\cite{Gheerardyn:2001jj} and they were found to have the form

\begin{equation}
  \begin{array}{rcl}
{\textstyle\frac{1}{2}}\delta_{\hat{\epsilon}}\hat{\psi}_{\hat{\mu}} 
 & = & 
\left\{
\hat{\nabla}_{\hat{\mu}}(\hat{\omega}+\hat{K})
\,+\,\textstyle{\frac{i}{288}}\left[
\hat{\Gamma}^{\hat{a}\hat{b}\hat{c}\hat{d}}{}_{\hat{\mu}}
\,-\, 8 \hat{\Gamma}^{\hat{b}\hat{c}\hat{d}}{\hat{e}_{\hat{\mu}}}^{\hat{a}}
\right] \hat{G}_{\hat{a}\hat{b}\hat{c}\hat{d}}
\right.\\
& & \\
& & 
\left.
-\textstyle{\frac{i}{12}}
     \hat{k}_{(n)\hat{\nu}}\mathsf{Q}^{nm}\hat{k}_{(m)}^{\hat{\nu}}
        \hat{\Gamma}_{\hat{\mu}}
+ \textstyle{\frac{i}{2}} \hat{k}_{(n)\hat{\mu}}\mathsf{Q}^{nm}
       \hat{k}_{(m)\hat{\nu}}\hat{\Gamma}^{\hat{\nu}}
\right\}
\hat{\epsilon} \, . \\
\end{array}
\end{equation}

It should be clear from this discussion that a fully supersymmetric
theory is obtained for each value of $n$. In the next sections we are
going to see how known gauged/massive supergravities arise in the
dimensional reduction of the above Lagrangian and supersymmetry
transformation rules in the direction of the Killing vectors
$\hat{k}_{(m)}$.

\section{Romans' Massive $N=2A,d=10$ Supergravity from $d=11$}
\label{sec-Romans}

The reduction of the $n=1$ case in the direction of the unique Killing
vector present with the same Kaluza-Klein Ansatz as in the massless
case gives Romans' massive $N=2A,d=10$ supergravity
\cite{Bergshoeff:1997ak} with the field content (in stringy notation)

\begin{equation}
\{g_{\mu\nu},\phi,B_{\mu\nu},C^{(3)}{}_{\mu\nu\rho},C^{(1)}{}_{\mu},
\psi_{\mu},\lambda\}  
\end{equation}

\noindent 
and with a mass parameter\footnote{This is the parameter $m_{R}$ of
  Refs.~\cite{Bergshoeff:2002mb,Bergshoeff:2002nv}.} $m$ equal to
minus the mass matrix $m=-\mathsf{Q}$.  Thus, setting for the bosonic
fields

\begin{equation}
\left( \hat{e}_{\hat{\mu}}{}^{\hat{a}} \right) =
\left(
\begin{array}{cc}
e^{-\frac{1}{3}\phi} e_{\mu}{}^{a}
&
e^{\frac{2}{3}\phi} C^{(1)}{}_{\mu}
\\
&
\\
0
&
e^{\frac{2}{3}\phi}
\\
\end{array}
\right) \, , 
\hspace{1cm}
\left( \hat{e}_{\hat{a}}{}^{\hat{\mu}} \right) =
\left(
\begin{array}{cc}
e^{\frac{1}{3}\phi} e_{a}{}^{\mu}
&
-e^{\frac{1}{3}\phi} C^{(1)}{}_{a}
\\
&
\\
0
&
e^{-\frac{2}{3}\phi}
\\
\end{array}
\right)\, ,
\label{eq:basis}
\end{equation}

\begin{equation}
\hat{C}_{\mu\nu\rho}=C^{(3)}{}_{\mu\nu\rho}\, ,
\hspace{1cm}
\hat{C}_{\mu\nu\underline{z}}=B_{\mu\nu}\, ,
\end{equation}

\noindent
 we get the string-frame bosonic action

\begin{equation}
\begin{array}{rcl}
S & = &
{\displaystyle\int} d^{10}x\,
\sqrt{|g|} \left\{ e^{-2\phi}
\left[ R -4\left( \partial\phi \right)^{2}
+{\textstyle\frac{1}{2\cdot 3!}} H^{2}\right]
-\left[ 
{\textstyle\frac{1}{2}}m^{2}
+{\textstyle\frac{1}{2\cdot 2!}} m^{2}\left(G^{(2)}\right)^{2}
+{\textstyle\frac{1}{2\cdot 4!}}\left(G^{(4)}\right)^{2}
\right]
\right.
\\
& & \\
& & 
\hspace{2cm}
-{\textstyle\frac{1}{144}} \frac{1}{\sqrt{|g|}}\
\epsilon\left[\partial C^{(3)}\partial C^{(3)}B
+{\textstyle\frac{1}{2}}m\partial C^{(3)}BBB
+{\textstyle\frac{9}{80}}m^{2}BBBBB\right]
\biggr \}\, ,\\
\end{array}
\label{eq:massiveIIAaction2}
\end{equation}

\noindent
 where $G^{(2)}$ and  $G^{(4)}$ are the RR 2- and 4-form 
field strengths

\begin{equation}
G^{(2)}=2\partial C^{(1)}+mB\, ,
\hspace{1cm}
G^{(4)}  =  4\partial C^{(3)} 
-12\partial BC^{(1)}+3m BB\, ,
\end{equation}

\noindent
 and $H$ is the NSNS 3-form field strength

\begin{equation}
H=3\partial B\, .  
\end{equation}

These field strengths and the Lagrangian are invariant under the
bosonic gauge transformations

\begin{equation}
\delta B  =  2\partial\Lambda\, ,
\hspace{1cm}
\delta C^{(1)} = \partial\Lambda^{(0)}-m\Lambda\, , 
\hspace{1cm}
\delta C^{(3)} = 3\partial\Lambda^{(2)}-3mB\Lambda -H\Lambda^{(0)}\, , 
\end{equation}

\noindent
 where the gauge parameters 
$\Lambda^{(2)}_{\mu\nu},\Lambda_{\mu},\Lambda^{(0)}$ are related to
the 11-dimensional ones $\hat{\chi}_{\hat{\mu}\hat{\nu}}$ and
$\hat{\xi}^{\hat{\mu}}$ (the generator of infinitesimal g.c.t.'s) by

\begin{equation}
\hat{\chi}_{\mu\nu} = \Lambda^{(2)}_{\mu\nu}\, ,
\hspace{1cm}
\hat{\chi}_{\mu\underline{z}}
={\textstyle\frac{1}{m}}\hat{\lambda}_{\mu}
=\Lambda_{\mu}\, ,
\hspace{1cm}
\hat{\xi}^{\underline{z}}=
\Lambda^{(0)}\, .
\end{equation}

The massive gauge invariance of this theory does not lead to a gauged
supergravity just because the dimensional reduction of the massive
gauge parameter only gives a 1-form. On the other hand, it allows us
to gauge away the RR vector leaving in the action a mass term for the
NSNS 2-form. This is the simplest example of the St\"uckelberg
mechanism that will be at work in all the cases that we are going to
review here. The St\"uckelberg vectors will always be the ones coming
from the metric (here the RR vector). The gauge vector fields (if any)
will always come from the 3-form.

The Ansatz for the fermionic fields and supersymmetry parameter is

\begin{equation}
\hat{\epsilon}  = e^{-\frac{1}{6}\phi}\, 
\epsilon\, , 
\hspace{1cm}
\hat{\psi}_{a} =
e^{\frac{1}{6}\phi}
\left(2\psi_{a} -{\textstyle\frac{1}{3}}
\Gamma_{a} \lambda\right)\, ,
\hspace{1cm}
\hat{\psi}_{z} = 
{\textstyle\frac{2i}{3}} e^{\frac{1}{6}\phi}
\Gamma_{11} \lambda\, ,
\end{equation}

\noindent
 and leads to the supersymmetry transformation rules 

\begin{equation}
\label{eq:massIIAsusyrules}
\left\{
\begin{array}{rcl}
\delta_{\epsilon} \psi_{\mu} & = & 
\left\{  
\partial_{\mu} -\frac{1}{4} \left(\not\!\omega_{\mu} 
+\frac{1}{2}\Gamma_{11}\not\!\! H_{\mu}\right) 
\right\} \epsilon 
+\frac{i}{8} e^{\phi} \Sigma_{n=0}^{n=2} \frac{1}{(2n)!}
\not\! G^{(2n)} \Gamma_{\mu} 
\left( -\Gamma_{11} \right)^{n}\epsilon\, , \\
& & \\
\delta_{\epsilon}\lambda & = &   
\left[\not\!\partial\phi
+\frac{1}{2\cdot 3!}\Gamma_{11}\not\!\! H
\right]
\epsilon 
+ \frac{i}{4} e^{\phi} \sum_{n=0}^{n=2} \frac{5-2n}{(2n)!} 
\not\! G^{(2n)} \left(-\Gamma_{11} \right)^{n} 
\epsilon\, .\\
\end{array}
\right.
\end{equation}

\noindent
 where we have identified

\begin{equation}
G^{(0)}=m\, .  
\end{equation}

\section{Massive $N=2,d=9$ Supergravities from $d=11$}
\label{sec-d=9}

The reduction of the $n=2$ case in the direction of the two Killing
vectors present with the same Kaluza-Klein Ansatz as in the massless
case gives gauged/massive $N=2,d=9$ supergravities characterized by
the mass matrices $\mathsf{Q}^{mn}$
\cite{Meessen:1998qm,Cowdall:2000sq,Hull:2002wg,Bergshoeff:2002mb,
  Nishino:2002zi,Bergshoeff:2002nv}.  The field content of these
theories is

\begin{equation}
\{g_{\mu\nu},\varphi,L_{m}{}^{i},C_{\mu\nu\rho},
B_{m\, \mu\nu},V_{\mu},A^{m}{}_{\mu},
\psi^{i}_{\mu},\lambda^{i}\}\, .  
\end{equation}

\noindent
 The $L_{m}{}^{i}$ parametrize an $Sl(2,\mathbb{R})/SO(2)$ coset.
The field $V_{\mu}$ comes from the 11-dimensional 3-form
components $\hat{C}_{\mu mn}$ and will be a gauge field. Its presence is the
main new feature with respect to the $n=1$ case. The gauge group will depend
on the choice of mass matrix, as we are going to see.  As in all cases, there
will always be the same number of 2-forms $B_{\mu\nu\, m}$ and Kaluza-Klein
vectors $A^{m}{}_{\mu}$ that play the role of St\"uckelberg fields for the
2-forms.

Explicitly, the Kaluza-Klein Ansatz for the bosonic fields is

\begin{equation}
\left( \hat{e}_{\hat{\mu}}{}^{\hat{a}} \right) = 
\left(
\begin{array}{cr}
e^{-\frac{1}{3\sqrt{7}}\varphi} e_{\mu}{}^{a} & 
e^{\frac{\sqrt{7}}{6}\varphi}L_{m}{}^{i}A^{m}{}_{\mu} \\
&\\
0             & e^{\frac{\sqrt{7}}{6}\varphi} L_{m}{}^{i}  \\
\end{array}
\right)
\, , 
\hspace{1cm}
\left(\hat{e}_{\hat{a}}{}^{\hat{\mu}} \right) =
\left(
\begin{array}{cr}
e^{\frac{1}{3\sqrt{7}}\varphi}e_{a}{}^{\mu} & 
-e^{\frac{1}{3\sqrt{7}}\varphi}A^{m}{}_{a}   \\
& \\
0             &  e^{\frac{-\sqrt{7}}{6}\varphi} L_{i}{}^{m}  \\
\end{array}
\right)\, , 
\end{equation}

\noindent
 and\footnote{The definition of $C_{\mu\nu\rho}$ is not the 
  most naive $\hat{C}_{abc}\sim C_{abc}$ because in this case one is
  interested in recoverig exactly the theories obtained by
  non-geometrical Scherk-Schwarz reduction from $N=2B,d=10$
  supergravity \cite{Meessen:1998qm}.}

\begin{equation}
\left\{
\begin{array}{rcl}
\hat{C}_{\mu\nu\rho} & = & C_{\mu\nu\rho} 
-\frac{3}{2}A^{m}{}_{[\mu}B_{m|\, \nu\rho]} 
+3\eta_{mn}V_{[\mu}A^{m}{}_{\nu}A^{n}{}_{\rho]}\, ,\\
& & \\
\hat{C}_{\mu\nu m} & = & B_{m\, \mu\nu}
-2\eta_{mn} V_{[\mu}A^{n}{}_{\nu]}\, ,\\
& & \\
\hat{C}_{\mu mn} & = & \eta_{mn}V_{\mu}\, .\\
\end{array}
\right.
\end{equation}

The gauge parameter $\hat{\chi}_{\hat{\mu}\hat{\nu}}$ gives rise to a
scalar parameter $\sigma$, two vector parameters $\lambda_{m\, \mu}$
and a 2-form parameter $\chi_{\mu\nu}$:

\begin{equation}
\hat{\chi}_{\mu \nu}  =  \chi_{\mu\nu}\, ,
\hspace{1cm}
\hat{\chi}_{\mu m} = \lambda_{m\, \mu}\, ,
\hspace{1cm}
\hat{\chi}_{mn} = \eta_{mn}\sigma\, ,
\end{equation}

The gauge vector $V_{\mu}$ transforms under the group generated by the
single\footnote{Some of the gauged/massive $N=2,d=9$ theories
  presented in Ref.\cite{Bergshoeff:2002nv} have a 2-parameter
  non-Abelian guage group and, tehrefore, cannot be described in this
  framework even if we allowed for more general, non-symmetric mass
  matrices.} local parameter $\sigma(x)$

\begin{equation}
\delta_{\sigma} V_{\mu} = \partial_{\mu}\sigma\, .  
\end{equation}

To find which is the one-parameter gauge group we have to look at the
$\delta_{\sigma}$ transformations of the fields that carry
$Sl(2,\mathbb{R})$ indices $m,n$:
 
\begin{equation}
  \begin{array}{rcl}
\delta_{\sigma}L_{m}{}^{i} & = & 
-\sigma L_{n}{}^{i}\mathsf{m}^{n}{}_{m}\, ,\\
& & \\
\delta_{\sigma}A^{m}{}_{\mu} & = & 
\sigma \mathsf{m}^{m}{}_{n} A^{n}{}_{\mu}\, , \\
& & \\
\delta_{\sigma}B_{\mu\nu\, m} & = & 
-\sigma  B_{m\, \mu\nu}\mathsf{m}^{n}{}_{m}
+2\eta_{mn}\partial_{[\mu}\sigma A^{n}{}_{\nu]}\, ,\\
\end{array}
\end{equation}

\noindent
 that leave invariant all the field strengths except for that 
of $B_{m\, \mu\nu}$ that transforms covariantly. This tells us that
the gauge group of the 9-dimensional theory is the group generated by
the $2\times 2$ traceless matrix $\mathsf{m}^{m}{}_{n}=
-\mathsf{Q}^{mp}\eta_{pn}$, which is a generator of a subgroup of
$Sl(2,\mathbb{R})$.  By construction, it is the subgroup that
preserves the mass matrix $\mathsf{Q}^{mn}$: it transforms
according to

\begin{equation}
\mathsf{Q}^{\prime} = \Lambda\mathsf{Q}\Lambda^{T}\, ,
\hspace{1cm}
\Lambda = e^{\sigma\mathsf{m}}\, ,
\end{equation}

\noindent
 then the condition that it is preserved $\Lambda^{-1}\mathsf{Q} =
\mathsf{Q}\Lambda^{T}$ 
translates into

\begin{equation}
\mathsf{m} \mathsf{Q}= - \mathsf{Q}\mathsf{m}^{T}\, ,  
\end{equation}

\noindent
 which is trivially satisfied for $\mathsf{m}=-\mathsf{Q}\eta$
on account of the property $\mathsf{m}^{T}= -\eta\mathsf{m}\eta^{-1}$.

It is clear that the theories obtained can be classified first by the sign of
the determinant of the mass matrix $\alpha^{2}= -4{\rm det}\,\mathsf{Q}$ ,
which is an $SL(2,\mathbb{R})$ invariant: class~I $\alpha^{2}=0$, class~II
$\alpha^{2}>0$ and class~III $\alpha^{2}<0$
\cite{Hull:2002wg,Bergshoeff:2002mb}. These classes should be subdivided
further into $Sl(2,\mathbb{Z})$ equivalence classes since the theories are
equivalent only when they are related by $Sl(2,\mathbb{Z})$ transformations.
However, it should be clear that theories within the same $\alpha^{2}$ class
have the same gauge group, the difference being a change of basis which is an
$Sl(2,\mathbb{R})$ but not an $Sl(2,\mathbb{Z})$ transformation.

Thus, all theories in class~III ($\alpha^{2}<0$) have gauge group $SO(2)$ and
all theories in class~II ($\alpha^{2}>0$) have gauge group $SO(1,1)$. The
theories in class~I ($\alpha^{2}=0$) are all equivalent to one with 

\begin{equation}
\mathsf{Q}=
\left(
  \begin{array}{cc}
-m & 0 \\
0 & 0 \\
  \end{array}
\right)\, , 
\end{equation}
 
\noindent
 which is just the reduction of the $n=1$ case (Romans' theory) 
considered in the previous section. The group is now $SO(1,1)$, with 

\begin{equation}
\Lambda =   
\left(
  \begin{array}{cc}
1 & \sigma m \\
0 & 1 \\
  \end{array}
\right)\, . 
\end{equation}

\noindent
The transformation laws of the fields of this theory are rather
unconventional but the theory is still a gauged supergravity.

From a combination of different terms we get the scalar potential 

\begin{equation}
{\cal V}\left(\varphi,{\cal M}\right) ={\textstyle\frac{1}{2}}
e^{\frac{4}{\sqrt{7}}\varphi}
{\rm Tr} \left(\mathsf{m}^{2} 
+\mathsf{m} {\cal M}\mathsf{m}^{T}{\cal M}^{-1}\right)\, ,
\hspace{1cm}
\mathcal{M}_{mn}= L_{m}{}^{i}L_{n}{}^{i}\, .
\end{equation}

\noindent
 Its presence suggests the existence of domain-wall (7-brane)
solutions which will be the vacua of the different theories obtained from
different mass matrices. In fact, these domain-wall solutions correspond to
different 7-brane solutions of the $N=2B,d=10$ theory: each kind of
10-dimensional 7-brane is characterized by its $Sl(2,\mathbb{Z})$ monodromy
$\Lambda$ and it is possible to reduce the $N=2B,d=10$ theory in the
Scherk-Schwarz generalized fashion admitting this monodromy for the different
fields. The result is a gauged/massive $N=2,d=9$ supergravity with a mass
matrix $\mathsf{Q}$ related to $\Lambda=e^{2\pi R \mathsf{m}}$ as explained
above. The domain-wall solutions and their 10-dimensional origin and
monodromies have been studied in detail in Ref.~\cite{Bergshoeff:2002mb}.  A
study of the non-conformal 8-dimensional field theories living in the
``boundaries'' of these solutions and their relations is still lacking.

The 2-forms $B_{m\, \mu\nu}$ are also invariant under the standard 2-form
gauge transformations

\begin{equation}
\delta_{\lambda} B_{m\, \mu\nu} = 2\partial_{[\mu}\lambda_{|m|\nu]}\, .
\end{equation}

\noindent
 This is possible because these transformations are supplemented by
the massive gauge transformations of the KK vectors

\begin{equation}
\delta_{\lambda} A^{m}{}_{\mu}= \mathsf{Q}^{mn}\lambda_{n\, \mu}\, ,
\end{equation}

\noindent
 that leave invariant the field strength

\begin{equation}
F^{m}{}_{\mu\nu}=2\partial_{[\mu}A^{m}{}_{\nu]}
-\mathsf{Q}^{mn}B_{n\, \mu\nu}\, ,  
\end{equation}

\noindent
 which would allow us to gauge them away  giving explicit 
mass terms to the 2-forms. It is in this way (St\"uckelberg mechanism)
that there is no clash between the gauge invariance under
$\delta_{\sigma}$ and the 2-form gauge transformations
$\delta_{\lambda}$.

The full bosonic action for this theory can be found in
Ref.~\cite{Meessen:1998qm} and the fermionic supersymmetry transformation
rules in Ref.~\cite{Gheerardyn:2001jj}. Their are not essential to our quick
review and we will not write them explicitly.

\section{Massive $N=4,d=8$ Supergravities from $d=11$}
\label{sec-d=8}

The reduction of next case $n=3$ in the direction of the three Killing
vectors gives 8-dimensional gauged theories
\cite{Alonso-Alberca:2000gh}. Only the $SO(3)$ case was studied in
Ref.~\cite{Alonso-Alberca:2000gh} but we are going to show that more
general non-compact and non-semisimple gaugings naturally arise as in
the previous case.  We are going to use the general formalism and
field definitions that will be valid in any dimension to show that in
the general case $n$ one can get $SO(n-l,l)$-gauged
$(11-n)$-dimensional supergravities.

The field content of these theories is

\begin{equation}
\{g_{\mu\nu},\varphi,a,L_{m}{}^{i},C_{\mu\nu\rho},
B_{m\, \mu\nu},V_{mn\, \mu},A^{m}{}_{\mu},
\psi^{i}_{\mu},\lambda^{i}\}\, ,  
\end{equation}

\noindent
 where the  indices $m,n=1,2,3$ are $Sl(3,\mathbb{R})$ indices
and also, simultaneously, gauge indices.  The $L_{m}{}^{i}$
parametrize now an $Sl(3,\mathbb{R})/SO(3)$ coset and the three vector
fields $V_{mn\, \mu}$ gauge a 3-dimensional group which should be a
subgroup of $Sl(3,\mathbb{R})$\footnote{In the general case we will
  have $Sl(n,\mathbb{R})$ indices, the $L_{m}{}^{i}$ will parametrize
  an $Sl(n,\mathbb{R})/SO(n)$ coset and instead of one scalar $a$
  one gets $a_{mnp}$ (here $a_{mnp}=a\epsilon_{mnp}$). Further, we will
  have $n(n-1)/2$ gauge vectors $V_{mn\, \mu}$.}.

The Anstaz for the bosonic fields is\footnote{In the general case only
  the powers of $\varphi$ are different.}

\begin{equation}
\left( \hat{e}_{\hat{\mu}}{}^{\hat{a}} \right) = 
\left(
\begin{array}{cr}
e^{\frac{1}{6}\varphi} e_{\mu}{}^{a} & 
e^{-\frac{1}{3}\varphi}L_{m}{}^{i}A^{m}{}_{\mu} \\
&\\
0             & e^{-\frac{1}{3}\varphi} L_{m}{}^{i}  \\
\end{array}
\right)
\, , 
\hspace{1cm}
\left(\hat{e}_{\hat{a}}{}^{\hat{\mu}} \right) =
\left(
\begin{array}{cr}
e^{-\frac{1}{6}\varphi}e_{a}{}^{\mu} & 
-e^{-\frac{1}{6}\varphi}A^{m}{}_{a}   \\
& \\
0             &  e^{\frac{1}{3}\varphi} L_{i}{}^{m}  \\
\end{array}
\right)\, , 
\end{equation}

\noindent
 and, using now the standard decomposition\footnote{In 
  the general case $\hat{C}_{ijk} \sim
  L_{i}{}^{m}L_{j}{}^{n}L_{k}{}^{p}a_{mnp}$.}

\begin{equation}
  \begin{array}{rclrcl}
\hat{C}_{abc} & = & e^{-\frac{1}{2}\varphi}C_{abc}\, ,&
\hat{C}_{abi} & = & L_{i}{}^{m} 
B_{m\, ab}\, ,\\
& & & & & \\
\hat{C}_{aij} & = & 
e^{\frac{1}{2}\varphi}L_{i}{}^{m} L_{j}{}^{n}V_{mn\, a}\, ,
\hspace{1.5cm}&
\hat{C}_{ijk} & = & e^{\varphi}\epsilon_{ijk} a\, ,\\
  \end{array}
\label{eq:d-8-potential-flat}
\end{equation}

\noindent
 as we are going to do from now on, we get\footnote{In 
  the general case we only have to substitute $a\epsilon_{mnp}$ by
  $a_{mnp}$.}

\begin{equation}
  \begin{array}{rcl}
\hat{C}_{\mu\nu\rho} & = & C_{\mu\nu\rho} +3A^{m}{}_{[\mu|}B_{m|\, \nu\rho]}
+3V_{mn\, [\mu}A^{m}_{\nu}A^{n}_{\rho]}
+a\epsilon_{mnp}A^{m}{}_{[\mu}A^{n}{}_{\nu}A^{p}{}_{\rho]}\, ,\\
& & \\
\hat{C}_{\mu\nu m} & = & B_{m\, \mu\nu} -2 V_{mn\, [\mu}A^{n}{}_{\nu]}
+a \epsilon_{mnp}A^{n}{}_{\mu}A^{p}{}_{\nu}\, ,\\
& & \\
\hat{C}_{\mu mn} & = & V_{mn\, \mu} +a\epsilon_{mnp}A^{p}{}_{\mu}\, ,\\
& & \\
\hat{C}_{mnp} & = & \epsilon_{mnp} a\, .\\
  \end{array}
\label{eq:d8-C-curved}
\end{equation}

The standard decomposition of the 4-form field strength 

\begin{equation}
\begin{array}{rclrcl}
\hat{G}_{abcd} & = & e^{-\frac{2}{3}\varphi} G_{abcd}\, ,&
\hat{G}_{abc\, i} & = & 
e^{-\frac{1}{6}\varphi} L_{i}{}^{m}\, H_{m\, abc}\, , \\
& & & & & \\
\hat{G}_{ab\, ij} & = & 
e^{\frac{1}{3}\varphi}\, L_{i}{}^{m}L_{j}{}^{n}
\left[F_{mn\, ab} + a \epsilon_{mnp}\, F^{p}{}_{ab} \right]\, , 
\hspace{1cm} &
\hat{G}_{a\, ijk} & = & 
e^{\frac{5}{6}\varphi}\, \epsilon_{ijk}\, \partial_{a}\, a\, ,\\
\end{array}
\end{equation}

\noindent
 gives the following field strengths

\begin{equation}
  \begin{array}{rcl}
&&
   G_{\mu\nu\rho\sigma} = 4\partial_{[\mu}C_{\nu\rho\sigma]}
   +6B_{m\, [\mu\nu}F^{m}{}_{\rho\sigma]} 
   -3B_{m\, [\mu\nu|}\mathsf{Q}^{mn}B_{n\, |\rho\sigma]}\, ,\\
 & & \\
&&
   H_{m\, \mu\nu\rho} = 3\partial_{[\mu|}B_{m|\, \nu\rho]}
   +3V_{mn\, [\mu}F^{n}{}_{\nu\rho]}\, ,\\
 & & \\
&&
   F_{mn\, \mu\nu} = 2\partial_{[\mu|}V_{mn|\, \nu]} 
   +2V_{mp\, [\mu|} \mathsf{Q}^{pq}V_{nq|\, \nu]}\, , \\
 & & \\
&&
   F^{m}{}_{\mu\nu} = 2\partial_{[\mu}A^{m}{}_{\nu]} 
   -\mathsf{Q}^{mn}B_{m\, \mu\nu}\, .\\
  \end{array}
\label{eq:d-8-massive-field-strengths}
\end{equation}

\noindent
 to which we have to add the covariant derivative of the 
$Sl(3,\mathbb{R})/SO(3)$ scalars\footnote{In the general case we also
  have to add the covariant derivative of the $a_{mnp}$ scalars
  \begin{equation}
\mathcal{D}_{\mu}a_{mnp} = \partial_{\mu} a_{mnp}
-3V_{[m|q\, \mu}\mathsf{Q}^{qr} a_{|np]r}\, ,
   \end{equation}
\noindent
 that reduces to a partial derivative in $d=8$ when
   $a_{mnp}=\epsilon_{mnp}a$.  
}

\begin{equation}
\mathcal{D}_{\mu}L_{m}{}^{i} = \partial_{\mu} L_{m}{}^{i}
-V_{mp\, \mu}\mathsf{Q}^{pq} L_{q}{}^{i}\, ,
\end{equation}

Let us now analyze the different gauge symmetries of the theory.  The
2-form $\hat{\chi}_{\hat{\mu}\hat{\nu}}$ decomposes now into a 2-form
$\chi_{\mu\nu}$, 3 vector parameters $\lambda_{m\, \mu}$ which will be
associated to the massive gauge invariances of the 3 2-forms $B_{m\,
  \mu\nu}$ and $3$ scalars $\sigma_{mn}=-\sigma_{nm}$\footnote{In the
  general case we will get $n$ vector parameters associated to the
  massive gauge invariances of the n 2-forms $B_{m\, \mu\nu}$ and
  $n(n-1)/2$ scalars $\sigma_{mn}=-\sigma_{nm}$.}

\begin{equation}
\hat{\chi}_{\mu \nu}  = \chi_{\mu\nu}\, ,
\hspace{1cm}
\hat{\chi}_{\mu m} = \lambda_{m\, \mu}\, ,
\hspace{1cm}
\hat{\chi}_{mn} = \sigma_{mn}\, .
\end{equation}

It is also convenient to define

\begin{equation}
\sigma^{m}{}_{n}=\mathsf{Q}^{mp}\sigma_{pn}\, .  
\end{equation}

These are going to be the infinitesimal generators of the gauge
transformations. Observe that, depending on the choice of
$\mathsf{Q}^{mp}$, $\sigma^{m}{}_{n}$ can contain an equal or smaller
number of independent components than $\sigma_{pn}$ and, thus, the
gauge group can have dimension $3$ or smaller.

Under the $\delta_{\sigma}$ transformations\footnote{In the general
  case the gauge transformations of these fields take the same form
  but the scalars $a_{mnp}$ transform covariantly
  \begin{equation}
\delta_{\sigma}a_{mnp}=-3a_{q[np}\sigma^{q}{}_{m]}\, .    
  \end{equation}
  This transformation vanishes in $d=8$ when
  $a_{mnp}=\epsilon_{mnp}a$.
}

\begin{equation}
  \begin{array}{rcl}
\delta_{\sigma}L_{m}{}^{i} & = & 
- L_{n}{}^{i}\sigma^{n}{}_{m}\, ,\\
& & \\
\delta_{\sigma}A^{m}{}_{\mu} & = & 
\sigma^{m}{}_{n} A^{n}{}_{\mu}\, , \\
& & \\
\delta_{\sigma}V_{mn\, \mu} &=& \mathcal{D}_{\mu}\sigma_{mn}\, ,\\
& & \\
\delta_{\sigma}B_{m\, \mu\nu} & = & 
-  B_{n\, \mu\nu}\sigma^{n}{}_{m}
+2\partial_{[\mu|}\sigma_{mn} A^{n}{}_{|\nu]}\, ,\\
& & \\
\delta_{\sigma} C_{\mu\nu\rho} & = & 
3\partial_{[\mu|}\sigma_{mn}A^{m}{}_{|\nu}A^{n}{}_{\rho]}\, ,\\
\end{array}
\label{eq:d-8-massive-gauge-transformations-1}
\end{equation}

\noindent
 the field strengths and covariant derivatives transform 
covariantly, i.e.

\begin{equation}
  \begin{array}{l}
\delta_{\sigma}G=0\, ,
\hspace{.5cm}  
\delta_{\sigma}H_{m} = -H_{n}\sigma^{n}{}_{m}\, ,
\hspace{.5cm}  
\delta_{\sigma}F_{mn} = -2F_{p[n}\sigma^{p}{}_{m]}\, ,
\hspace{.5cm}  
\delta_{\sigma}F^{m}=\sigma^{m}{}_{n}F^{n}\, ,\\
\\
\delta_{\sigma}\mathcal{D}L_{m}{}^{i}=
-(\mathcal{D}L_{n}{}^{i})\sigma^{n}{}_{m}\, ,
\hspace{.5cm}  
\delta_{\sigma}\mathcal{D}a_{mnp} =
-3(\mathcal{D}a_{q[np})\sigma^{q}{}_{m]}\, .\\
\end{array}
\end{equation}

The gauge group is the orthogonal subgroup of $Sl(3,\mathbb{R})$
($Sl(n,\mathbb{R})$ in the general case) obtained by exponentiation of
$\sigma^{m}{}_{n}$ that preserves the mass matrix $\mathsf{Q}^{mn}$,
i.e.

\begin{equation}
\Lambda\mathsf{Q}\Lambda^{T}=\mathsf{Q}\, ,
\hspace{1cm}
\Lambda= e^{\sigma}\, .  
\end{equation}

\noindent
It is easy to see that the infinitesimal form of the above condition

\begin{equation}
\sigma^{m}{}_{p}\mathsf{Q}^{pn} =-\sigma^{n}{}_{p}\mathsf{Q}^{mp}\, ,  
\end{equation}

\noindent
is trivially satisfied. The generators of the gauge group
in this representation are the matrices

\begin{equation}
\Gamma(M^{mn})^{p}{}_{q} = 2\mathsf{Q}^{m[p}\delta^{n]}{}_{q}\, ,
\,\,\,\,
\Rightarrow 
\sigma^{p}{}_{q}= {\textstyle\frac{1}{2}}\sigma_{mn}\Gamma(M^{mn})^{p}{}_{q}\, ,
\end{equation}

\noindent
and the algebra they satisfy can be easily computed 

\begin{equation}
[M^{pq},M^{rs}] ={\textstyle\frac{1}{2}}f^{pq\, rs}{}_{mn} M^{mn}\, ,
\hspace{1cm}
f^{pq\, rs}{}_{mn} = 
8\delta^{[p}{}_{[m}\mathsf{Q}^{q][r}\delta^{s]}{}_{n]}\, .
\end{equation}

\noindent
Actually, using the above structure constants $f^{pq\, rs}{}_{mn}$ the
gauge field strength $F_{mn}$ can be written in the standard form

\begin{equation}
F_{mn} =2\partial V_{mn} 
+{\textstyle\frac{1}{4}}f^{pq\, rs}{}_{mn}V_{pq}V_{rs}\, .  
\end{equation}

\noindent
It is clear that the gauge groups $SO(3)$ and $SO(2,1)$ correspond to
the non-singular diagonal mass matrices\footnote{It is also clear that
  in the general case all the orthogonal subgroups $SO(n-l,l)$ of
  $Sl(n,\mathbb{R})$ are also included.} $\mathsf{Q}= \pm g\, {\rm
  diag}(+++)$ and $\mathsf{Q}= \pm g\, {\rm diag}(++-)$ respectively, but other groups
can also appear. For $n=3$ we can easily classify all the gauge groups
that appear, comparing with the Bianchi classification of all real
3-dimensional Lie algebras\footnote{This study is more complicated for
  $n>3$ and, further, the real Lie algebras are not classifiedd in
  general, but only for $n=3$ (the well-known Bianchi classification)
  and $n=4$. See e.g.~Ref.~\cite{kn:KSMCH} and references therein.}.

It is useful to change the notation first:

\begin{equation}
\label{eq:structureconstants}
V^{m}{}_{\mu} \equiv 
{\textstyle\frac{1}{2}}\epsilon^{mnp}V_{np\, \mu}\, ,
\hspace{.5cm}
T_{m}\equiv {\textstyle\frac{1}{2}}\epsilon_{mnp}M^{np}\, ,  
\hspace{.5cm}
f_{uv}{}^{w}\equiv {\textstyle\frac{1}{8}}
\epsilon_{upq}\epsilon_{vrs}\epsilon^{wmn}f^{pq\, rs}{}_{mn}
= -\epsilon_{uvt}\mathsf{Q}^{tw}\, .
\end{equation}

\noindent

The Bianchi classification starts with the observation that the most
general structure constants for a 3-dimensional Lie algebra can be
written in the form

\begin{equation}
\label{eq:generalstructtureconstants}
f_{ij}{}^{k} = -\epsilon_{ijl}b^{lk}  +2 \delta_{[i}{}^{k}a_{j]}\, ,
\hspace{1cm}
b^{lk}=b^{kl}\, ,
\hspace{1cm}
b^{kl}a_{l}=0\, .
\end{equation}

The next step consists in the diagonalization of the symmetric matrix
$b^{kl}$ whose eigenvalues are normalized to $\pm 1,0$ and the
determination of their possible null eigenvectors $a_{i}$. Comparing
with the structure constants Eq.~(\ref{eq:structureconstants}), we see
that we can obtain all the Lie algebras in the Bianchi classification
with $a_{i}=-\frac{1}{2}f_{ik}{}^{k}=0$. These are the Bianchi I, II,
VI, VII, VIII ($so(2,1)$) and IX ($so(3)$) algebras. 

How about the remaining Bianchi III, IV and V algebras? It is a simple
exercise to rewrite the general structure constants
Eq.~(\ref{eq:generalstructtureconstants}) in terms of just one
constrained matrix $d^{lk}=b^{kl} -\epsilon^{kli}a_{i}$ with no
definite symmetry properties:

\begin{equation}
f_{ij}{}^{k} = \epsilon_{ijl}d^{lk}\, ,
\hspace{1cm}
d^{(lk)}\epsilon_{kij}d^{ij}=0\, .
\end{equation}

\noindent 
This seems to suggest that we could cover all the possible cases by
allowing for a general, non-symmetric $Q^{mn}$, but this has to be
checked in detail.

Let us conclude the study of the gauge symmetries of the theory: the
parameters $\lambda_{m\, \mu}$ generate massive gauge transformations
under which

\begin{equation}
\delta_{\lambda}A^{m}{}_{\mu} = \mathsf{Q}^{mn}\lambda_{n\, \mu}\, , 
\hspace{1cm}
\delta_{\lambda}B_{m\, \mu\nu} = 2\partial_{[\mu|}\lambda_{m\, |\nu]}\, ,
\hspace{1cm}
\delta_{\lambda} C_{\mu\nu\rho} = 
-6A^{m}{}_{[\mu}\partial_{\nu|}\lambda_{m|\, \rho]}\, ,
\label{eq:d-8-massive-gauge-transformations-2}
\end{equation}

\noindent
 leaving invariant all the field strengths. In this and all 
cases this invariance can be used to eliminate the 3 ($n$) KK vector
fields $A^{m}{}_{\mu}$ giving masses to the 3 ($n$) 2-forms $B_{m\,
  \mu\nu}$.  The action for the full theory can be found in
Ref.~\cite{Alonso-Alberca:2000gh}.

Let us now compare the theory obtained, with
$\mathsf{Q}=g \mathbb{I}_{3\times 3}$ and gauge group $SO(3)$ (the other
cases cannot be compared) with Salam \& Sezgin's (SS)
\cite{Salam:1984ft}. The field contents are identical, only the
couplings are different: in the SS case the gauge vector fields are
the KK ones $A^{m}{}_{\mu}$ and the St\"uckelberg vector fields are
the ones coming from the 3-form $V^{m}{}_{\mu}$, while in our case
these roles are interchanged (the 2-forms are always massive). Some of
the couplings to the scalars $\varphi$ and $a$ are also different.

Actually it is convenient to describe the differences between both
8-dimensional theories through the action of the global $Sl(2,\mathbb{R})$
duality symmetry that the (equations of motion of the) massless theory enjoys.
The scalars $\varphi$ and $a$ can be combined in the complex scalar
$\tau=a+ie^{-\varphi}$ that parametrizes the coset $Sl(2,\mathbb{R})/SO(2)$
and undergoes fractional-linear transformations under $Sl(2,\mathbb{R})$.
Under this group, the vector fields form 3 doublets
$(V^{m}{}_{\mu},A^{m}{}_{\mu})$ while the 2-forms are
singlets\footnote{Actually, the 2-forms are singlets after a field
  redefinition.}. The 4-form field strength $G$ undergoes electric-magnetic
duality rotations.

The differences between the two 8-dimensional gauged theories are
associated, precisely, to the $Sl(2,\mathbb{R})$ transformation

\begin{equation}
S=
\left(
  \begin{array}{cc}
0 & 1 \\
-1 & 0\\
  \end{array}
\right)  
\end{equation}

\noindent
 that interchanges the vector fields $V^{m}{}_{\mu}$ and 
$A^{m}{}_{\mu}$ and transforms $\tau$ into $-1/\tau$. This is
reflected in the scalar potential which in our case is given by

\begin{equation}
{\cal V} =-{\textstyle\frac{1}{2}}
{\displaystyle\frac{|\tau|^{2}}{\Im{\rm m}(\tau)}}
\left\{ [{\rm Tr}\, (\mathsf{Q}{\cal M})]^2 
-2 {\rm Tr}\,(\mathsf{Q}{\cal M})^{2} \right\}\, ,
\end{equation}

\noindent
 while in SS's case is

\begin{equation}
{\cal V}_{\rm SS} =
-{\textstyle\frac{1}{2}}\, g^2
{\displaystyle\frac{1}{\Im{\rm m}(\tau)}}
\left\{({\rm Tr}\, \mathcal{M})^{2}
-2{\rm Tr}\, ( \mathcal{M}^{2} )
\right\} \, .
\end{equation}

Thus we can view our theory as the S-dual of SS's although, in
practice, one cannot perform such a transformation directly on the
gauged theories and, rather, one would have to do it in the ungauged
one. 

The non-compact gaugings that we obtain from ``massive 11-dimensional
supergravity'' have no known ``S-dual'', although it should be
possible to obtain them by the analytical continuation methods of
Ref.~\cite{Hull:2002cv}. Their 11-dimensional origin is unknown. As
for the non-semisimple gaugings, their ``S-duals'' are also unknown,
but now analytical continuation cannot be used to construct them.

\section{Massive $N=8,d=5$ Supergravities from $d=11$}
\label{sec-d=5}

From the discussions and examples in the previous sections it should
be clear that in the $n=4$ case we will obtain $SO(4-l,l)$-gauged
7-dimensional supergravities etc. A particularly interesting case is
the $n=6$ one, in which we can obtain $SO(6-l,l)$-gauged $N=8,d=5$
supergravities which were constructed in
Refs.~\cite{Gunaydin:1984qu,Pernici:ju}. This offers us the
possibility to check our construction and show that, as we have
claimed, it systematically gives gauged/massive supergravities.

The derivation of the 5-dimensional theory from ``massive 11-dimensional
supergravity'' offers no new technical problems and the action, field
strengths etc.~can be found applying the general recipes explained in the
previous section and are written explicitly in
Appendix~\ref{sec-reductiontod=5}. One of the highlights of this derivation
is the field content which is of the general form

\begin{equation}
\{g_{\mu\nu},\varphi,a,a_{mnp},L_{m}{}^{i},
B_{m\, \mu\nu},V_{mn\, \mu},A^{m}{}_{\mu},
\psi^{i}_{\mu},\lambda^{i}\}\, ,  
\end{equation}

\noindent
 where now the $m,n,p$ indices are  $Sl(6,\mathbb{R})$ indices
and where we have dualized the 3-form $C_{\mu\nu\rho}$ into the scalar $a$.
The scalars $\varphi$ and $a$ can be combined again into the complex $\tau$
that parametrizes $Sl(2,\mathbb{R})/SO(2)$. In the ungauged/massless theory
this $Sl(2,\mathbb{R})$ global symmetry and the more evident $Sl(6,\mathbb{R})$
are part of the $E_{6}$ duality group of the theory that only becomes manifest
after the 6 2-forms are also dualized into 6 additional vector
fields\footnote{The bosonic action of the massless theory with
  $C_{\mu\nu\rho}$ dualized into $a$ and the $B_{m\, \mu\nu}$ dualized into
  vector fields $N^{m}{}_{\mu}$ is given in Eq.~(\ref{eq:d5action-2}).}.

As usual, this is also the field content of the ungauged theory. This is
already a surprise since in Refs.~\cite{Gunaydin:1984qu,Pernici:ju} it was
argued that the theory could only be consistently gauged if the 6 KK vector
fields $A^{m}{}_{\mu}$ were dualized into 6 2-forms $\tilde{B}_{m\, \mu\nu}$
which, together with the already existing 6 2-forms $B_{m\, \mu\nu}$ and via a
self-dual construction, could describe 6 massive 2-forms. Once there are no
massless higher-rank fields with $Sl(6,\mathbb{R})$ indices left, the theory
can be consistently gauged. In the theory that we get, the same goal is
achieved by the St\"uckelberg mechanism: the 6 KK vector fields
$A^{m}{}_{\mu}$ are not dualized but are gauged away leaving mass terms for the
already existing 6 2-forms $B_{m\, \mu\nu}$.

Another interesting point is the form of the scalar potential
$\mathcal{V}(\varphi,a_{mnp})$, given in
Eq.~(\ref{eq:5-potential-1}). The first term, which is universal for
all the gauged/massive theories we are studying and is the only one
that survives the consistent truncation $a_{mnp}=0$, is independent of
the scalar $\varphi$ that measures the volume of the internal
manifold. As shown in Appendix~\ref{ap:potential-extrema}, this
universal term is always minimized for
$\mathcal{M}=\mathbb{I}_{n\times n}$ when $\mathsf{Q}=g\,
\mathbb{I}_{n\times n}$ and the value of the potential for $n=6$ is
constant and the vacuum is $AdS_{5}$ as in
Refs.~\cite{Gunaydin:1984qu,Pernici:ju}. Not only the vacuum is the
same: in Ref.~\cite{Gunaydin:1986cu} it was shown that there is is a
consistent truncation of the scalars that leaves the same potential
(the first term of Eq.~(\ref{eq:5-potential-1})) for the remaining
scalars and thus, in spite of the apparent differences it is plausible
that the two untruncated potentials are completely equivalent.

If the field content is equivalent, the symmetries of the theory are
the same the vacuum is the same and, presumably the potentials are
equivalent, we can expect to have obtained a completely equivalent
form of the $SO(6-l,l)$-gauged $N=8,d=5$ theories constructed in
Refs.~\cite{Gunaydin:1984qu,Pernici:ju}. To show or, rather, to make
more plausible this equivalence we would like to show that these
theories have identical equations of motion, but this is extremely
complicated for the full theories and we will content ourselves with
showing the equivalence of the self-dual and St\"uckelberg Lagrangians
for charged 2-forms ignoring all the scalars for the sake of
simplicity.

\subsection{Self-Duality versus St\"uckelberg}
\label{ap:self-duality}

The gauging of $N=4,d=7$ \cite{Pernici:zw} and $N=8,d=5$
\cite{Gunaydin:1984qu,Pernici:ju} supergravity theories presents many
peculiar features and problems absent in other cases. All these
problems were resolved using the {\sl self-duality mechanism}
\cite{Townsend:ad,Townsend:xs}.  Before comparing it with the
St\"uckelberg mechanism, we will review the above mentioned problems
and the reasoning that lead to the use of the self-duality mechanism
to solve them in the 5-dimensional case.

In the usual gauging procedure one gauges the symmetry group of all
the vector fields present in the ungauged theory. In one version of
$N=8,d=5$ ungauged supergravity in which all 2-forms have been
dualized into vectors, there are 27 vector fields, but there is no
27-dimensional simple Lie group, and therefore the standard recipe
does not work. The origin of the gauged theory from IIB
supergravity compactified on $S^{5}$
\cite{Gunaydin:1984fk,Kim:1985ez,Cvetic:2000nc} suggested the gauging
of the isometry group of the internal space, the 15-dimensional
$SO(6)$. $E_{6(6)}$ being the global symmetry group of the ungauged
theory and $Usp(8)$ the local composite one, the idea was to gauge an
$SO(6)$ subgroup of the $Usp(8)$ embedded in
$E_{6(6)}$\footnote{$Usp(8)$ contains $Sl(2,\mathbb{R})\times
  Sl(6,\mathbb{R})$ as a subgroup, and the $SO(6)$ to be gauged is in
  $Sl(6,\mathbb{R})$. One may also gauge a non-compact group
  $SO(6-l,l)$ instead of $SO(6)$.}. All bosonic fields are in
irreducible representations of $E_{6(6)}$ and in general transform as
reducible representations under $SO(6)$. In particular, the
$\mathbf{27}$ of vector fields breaks, under $SO(6)$, as
$\mathbf{27}=\mathbf{15}+\mathbf{6}+\mathbf{6}$. The $\mathbf{15}$ is
precisely the adjoint of $SO(6)$. This raises a second problem: how to
couple the two sextets of abelian vector fields to the 15 $SO(6)$
Yang-Mills fields.

On the other hand, the superalgebra of the gauged theory was expected
to be $SU(2,2|4)$. The irreducible representation of this superalgebra
in which the graviton is contained also contains two sextets of
2-index antisymmetric tensor fields (2-forms). This and other reasons
\cite{Gunaydin:1984fk,Kim:1985ez} suggested the replacement of the two
sextets of abelian vector fields by two sextets of 2-form fields, but
there is also a problem in coupling these fields to the Yang-Mills
ones: it is not possible to reconcile both gauge invariances
simultaneously.  Replacing ordinary derivatives by Yang-Mills
covariant ones breaks the local gauge invariance of the antisymmetric
fields, which means that there are more modes propagating than in the
ungauged theory. But there is a way out: the antisymmetric fields must
satisfy self-dual equations of motion (to be described later).

Once the twelve vectors have been replaced by the self-dual 2-form
fields one finds that the latter do not satisfy Bianchi identities,
and for consistency the model must be gauged
\cite{Gunaydin:1984qu,Pernici:ju}. This, in turn, implies that,
naively, the gauged theory does not have a good $g\rightarrow 0$
limit, although the limit can be taken after elimination of one of the
2-form sextets \cite{Townsend:ad}. In our (St\"uckelberg) formulation,
the $g\rightarrow 0$ limit can always be taken.

In the next two subsections we are going to construct St\"uckelberg
formulations for a massive, uncharged 2-form field and for a sextet of
massive 2-form fields charged under $SO(6)$ Yang-Mills fields and
we will show that they lead to equations of motion fully equivalent to
those obtained from self-dual formulations. The St\"uckelberg
formulations are just simplifications of our gauged/massive $N=8,d=5$
supergravity theory.

\subsubsection{Uncharged case}

We start from the standard action for a massive 2-form field

\begin{equation}
 S[B] = \int d^{5}x \left\{ {\textstyle\frac{1}{2\cdot 3!}} H^2 
- {\textstyle\frac{1}{4}} m^2 B^2 \right\}\, ,
\label{eq:action1-1}
\end{equation}

\noindent
where $H=3\partial B$.  The equation of motion for $B$ derived from
(\ref{eq:action1-1}) is the Proca equation

\begin{equation}
 (\Box + m^2) B_{\mu\nu} =0\, .
\end{equation}

The action given in (\ref{eq:action1-1}) is not gauge invariant.
To recover formally gauge invariance we introduce in the action
a St\"uckelberg field $A_{\mu}$, such that the action is now

\begin{equation}
 S[B,A] = \int d^{5}x \left\{ {\textstyle\frac{1}{2\cdot 3!}} H^2 
-{\textstyle\frac{1}{4}} F^2 \right\}\, ,
\label{eq:action1-2}
\end{equation}

\noindent
where

\begin{equation}
\begin{array}{rcl}
&&  H=3\partial B\, ,\\
 & & \\
&&  F = 2\partial A - mB\, .\\
\end{array}
\end{equation}

The equations of motion for these fields are

\begin{equation}
\begin{array}{rcl}
&&  \partial_{\mu} H^{\mu\nu\rho} - m F^{\nu\rho} = 0\, ,\\
 & & \\
&&  \partial_{\mu} F^{\mu\nu}=0\, ,\\
\end{array}
\end{equation}

\noindent
and now we have invariance under the following ``massive gauge
transformations":

\begin{equation}
\begin{array}{rcl}
&&  \delta A = m \Lambda\, ,\\
 & & \\
&&  \delta B = 2\, \partial \Lambda\, .\\
\end{array}
\end{equation}

The vector $A_{\mu}$ does not propagate any degrees of freedom, since
it can be completely gauged away.  In fact, setting $A_{\mu}=0$ we
recover the Proca equation.  So, as we know, the introduction of the
St\"uckelberg field is just a way of re-writing the theory described
by (\ref{eq:action1-1}) in a formally gauge invariant way.

Now, to connect with the self-dual formulation, we dualize the vector
$A_{\mu}$ into a two-form $\tilde{B}_{\mu\nu}$: we add a Lagrange
multiplier term in the action:

\begin{equation}
 S[B,\tilde{B},F] = \int d^{5}x
                      \left\{
{\textstyle\frac{1}{2\cdot 3!}} H^2 
-{\textstyle\frac{1}{4}} m^2 B^2
+{\textstyle\frac{1}{4}}\epsilon \partial\tilde{B} (F+mB)
                      \right\}\, .
\label{eq:action1-3}
\end{equation}

\noindent
The equation of motion for $F=dA$ is

\begin{equation}
 F = {}^{*}\tilde{H} = {\textstyle\frac{1}{3!}}\epsilon \tilde{H},
\label{eq:dual-F-1}
\end{equation}

\noindent
where $\tilde{H}=3\partial\tilde{B}$.
Inserting Eq.~(\ref{eq:dual-F-1}) into (\ref{eq:action1-3}) one gets

\begin{equation}
 S[B,\tilde{B}] = \int d^{5}x
                      \left\{
{\textstyle\frac{1}{2\cdot 3!}} H^2
+{\textstyle\frac{1}{2\cdot 3!}} \tilde{H}^2
+{\textstyle\frac{m}{12}} \epsilon \tilde{H} B
                      \right\}\, .
\label{eq:action1-4}
\end{equation}

The action above contains two 2-forms, but it describes the degrees of
freedom of only one massive 2-form.  Observe that this action is
invariant under the gauge transformations

\begin{equation}
\delta B = 2 \partial \Sigma\, ,
\hspace{1cm}
\delta \tilde{B} = 2 \partial \tilde{\Sigma}\, . 
\end{equation}

Using this gauge invariance, the equations of motion derived from
(\ref{eq:action1-4}) can always be integrated to yield\footnote{These
  two equations can be combined to get the Proca equation.}

\begin{equation}
\begin{array}{rcl}
&&  {}^{*}H = + m \tilde{B}\, ,\\
 & & \\
&&  {}^{*}\tilde{H} = - m B\, ,\\
\end{array}
\label{eq:eoms-LSD-uncharged}
\end{equation}

\noindent
which are precisely the equations of motion that one can derive
directly from the {\it self-dual} action:

\begin{equation}
 S_{SD}[B,\tilde{B}] = \int d^{5}x
                      \left\{
-{\textstyle\frac{1}{4}} m^2 B^2
-{\textstyle\frac{1}{4}} m^2 \tilde{B}^2
-{\textstyle\frac{m}{12}} \epsilon \tilde{H} B
                      \right\}\, .
\label{eq:LSD-1}
\end{equation}

Therefore, the self-dual action Eq.~(\ref{eq:LSD-1}) and
(\ref{eq:action1-4}) (and, therefore, the St\"uckelberg action
Eq.~(\ref{eq:action1-2})) are classically equivalent actions since
they lead to the same equations of motion.

Our next step 
will be to establish a relation between the St\"uckelberg and
self-dual actions for a sextet of $SO(6)$-charged, massive 2-forms.

\subsubsection{The $SO(6)$ Charged Case}

Let us consider now six massive two forms coupled to the 15
$SO(6)$-vector fields $V_{mn}$.  The St\"uckleberg action for them can
be read off from the action describing the 5-dimensional
massive/gauged supergravity, given explicitely in
Appendix~(\ref{sec-reductiontod=5}) setting
$\mathsf{Q}^{mn}=m\delta^{mn}$.
To simplify matters we truncate all
the fields that are not relevant for our problem (in particular, all
the scalars) and will work in flat spacetime. We are left with

\begin{equation}
 S[B_m,A_m,V_{mn}] = \int d^{5}x
            \left\{
{\textstyle\frac{1}{2\cdot 3!}} \mathbb{H}_{m}\mathbb{H}_{m}
-{\textstyle\frac{1}{4}} F_{m}F_{m}
-{\textstyle\frac{1}{4}} {\cal F}_{m}{\cal F}_{m}
            \right\}\, ,
\label{eq:action2-2}
\end{equation}

\noindent
where

\begin{equation}
\begin{array}{rcl}
&&  \mathbb{H}_{m} = 
3\partial B_{m} + 3V_{mn}F_{n}  \equiv H_{m} + 3V_{mn}F_{n}\, ,\\
 & & \\
&&  F_{m} = 2\partial A_{m} - m B_{m}\, ,\\
 & & \\
&&  {\cal F}_{mn} = 2\partial V_{mn} + 2 m V_{mp}V_{np}\, .\\
 & & \\
\end{array}
\label{eq:field-strengths-2}
\end{equation}

\noindent
where ${\cal D}$ is the $SO(6)$ covariant derivative.

This action is invariant under 

\begin{equation}
\begin{array}{rcl}
&& \delta A_{m} = \sigma_{mn} A_{n} + m \lambda_{m}\, ,\\
& & \\
&& \delta V_{mn} = {\cal D}\sigma_{mn}\, ,\\
& & \\
&& \delta B_{m} = 2\partial\lambda_{m}
     + 2\partial \sigma_{mn}A_{n} + m \sigma_{mn}B_{n}\, ,\\
\end{array}
\end{equation}

In order to dualize the vectors $A_{m}$ into two-forms $\tilde{B}_{m}$
we follow exactly the same steps as in the uncharged case, and the
(much more complicated) equation we find for $F_{m}$ is

\begin{equation}
 F_{m}{}^{\mu\nu} = 
{{\cal P}^{-1} (V)}_{mn}{}^{\mu\nu}{}_{\rho\sigma}
              \left[
                ({}^{*}\tilde{H})_{n}{}^{\rho\sigma}
                + H_{p}{}^{\rho\sigma\lambda}V_{pn\,\lambda}
              \right]\, ,
\label{eq:dual-F-2}
\end{equation}

\noindent
where

\begin{equation}
{\cal P}_{mn}{}^{\rho\sigma}{}_{\mu\nu} (V) =
                        \delta_{mn}\, \eta^{[\rho\sigma]}{}_{\mu\nu}
                        - 3 \eta^{[\rho\sigma}{}_{\mu\nu}
                          V_{np}{}^{\lambda]} V_{mp\, \lambda}\, ,
\end{equation}

Then, the action in terms of the dual fields $\tilde{B}_{m}$ reads

\begin{equation}
\begin{array}{rcl}
 S[B_{m},\tilde{B}_{m},V_{mn}] &=& {\displaystyle\int} d^{5}x
                      \left\{
{\textstyle\frac{1}{2\cdot 3!}} H_{m}H_{m}
+{\textstyle\frac{1}{4}} ({}^{*}\tilde{H}_{m} + H_{p}V_{pm})
{\cal P}^{-1}_{mn}
({}^{*}\tilde{H}_{n} + H_{q}V_{qn})
                      \right.\, \\
 & & \\
                     &&
                      \left.
-{\textstyle\frac{1}{4}}{\cal F}_{mn}{\cal F}_{mn}
+{\textstyle\frac{1}{12}} \epsilon \tilde{H}_{m} B_{m}
                      \right\}\, .\\
\end{array}
\label{eq:action2-4}
\end{equation}

The action given in (\ref{eq:action2-4}) describes only the degrees of
freedom of the six massive 2-forms $B_{m}$ coupled to the vector
fields $V_{mn}$.  This action is invariant under the following gauge
transformations

\begin{equation}
\begin{array}{rcl}
&& \delta V_{mn} = {\cal D}\sigma_{mn}\, ,\\
& & \\
&& \delta B_{m} = {\cal P}^{-1}_{mn}
                  \left\{
                    \left(
                      d\lambda_{n} - \frac{1}{2} \epsilon\, d\tilde{\lambda}_{p} V_{np}
                    \right)
                    - \frac{1}{2} \epsilon
                    \left[
                      \left( \tilde{B}_{p} - \frac{1}{2} \epsilon B_{q} V_{pq} \right) {\cal D} \sigma_{np}
                    \right]
                  \right\} \, ,\\
& & \\
&& \delta \tilde{B}_{m} = {\cal P}^{-1}_{mn}
                  \left\{
                    \left(
                      d\tilde{\lambda}_{n} - \frac{1}{2} \epsilon\, d\lambda_{p} V_{np}
                    \right)
                    - \frac{1}{2} \epsilon
                    \left[
                      \left( B_{p} - \frac{1}{2} \epsilon \tilde{B}_{q} V_{pq} \right) {\cal D} \sigma_{np}
                    \right]
                  \right\} \, .\\
\end{array}
\end{equation}

The equations of motion derived from (\ref{eq:action2-4}) are

\begin{equation}
\begin{array}{rcl}
&&  \mathcal{D}_{\mu} {\cal F}_{mn}{}^{\mu\nu}
             = \frac{1}{4} m^2  
               \epsilon^{\nu\rho\sigma\delta\lambda} 
B_{[m|\, \rho\sigma} \tilde{B}_{|n]\, \delta\lambda}\, ,\\
 & & \\
&&  {}^{*}H_{m} =
       + m [\tilde{B}_{m} + \frac{1}{2} \epsilon V_{mn} B_{n}]\, ,\\
 & & \\
&&  {}^{*}\tilde{H}_{m} =
       - m [B_{m} + \frac{1}{2}\epsilon V_{mn}  \tilde{B}_{n}]\, ,\\
\end{array}
\end{equation}

\noindent
which can also be derived from the {\it self-dual} action:

\begin{equation}
 S_{SD}[\vec{B}_{m},V_{mn}] = \int d^{5}x
                      \left\{
-{\textstyle\frac{1}{4}} m^2 \vec{B}^{T}_{m} \vec{B}_{m}
-{\textstyle\frac{1}{4}}{\cal F}_{mn}{\cal F}_{mn}
-{\textstyle\frac{m}{8}} \epsilon
\vec{B}^{T}_{m}  \eta \mathcal{D} \vec{B}_{m}\,
                      \right\}\, ,
\label{eq:LSD-2'}
\end{equation}

\noindent
where

\begin{equation}
 \vec{B}_{m}=
\left(
\begin{array}{cc}
 B_{m} \\
 \tilde{B}_{m} \\
\end{array}
\right)\, ,
\hspace{1cm}
\eta =
\left(
\begin{array}{cc}
 0 & 1 \\
 -1 & 0 \\
\end{array}
\right)\, ,
\end{equation}

\noindent
and $\mathcal{D}$ is the $SO(6)$ covariant derivative acting on
$\vec{B}_{m}$:

\begin{equation}
 \mathcal{D}\vec{B}_{m}=
\left(
\begin{array}{cc}
 \partial B_{m} - m V_{mn} B_{n} \\
\\
 \partial \tilde{B}_{m} + m V_{mn}\tilde{B}_{n}\\
\end{array}
\right)\, .
\label{eq:gauge-derivative}
\end{equation}

Observe that the $SO(6)$ charges of $B_{m}$ and $\tilde{B}_{m}$ are opposite.

This is precisely the kind of action that appears in the standard form
of $N=8,d=5$ gauged supergravity.

\section*{Acknowledgements}

N.A.-A. would like to thank P.~Meessen and P.~Resco for interesting
conversations, and specially E.~Lozano-Tellechea. T.O.~would like to
thank E.~Bergshoeff for interesting conversations and and specially
P.K.~Townsend for pointing us to Ref.~\cite{Townsend:ad}, the Newton
Institute for Mathematical Sciences and the Institute for Theoretical
Physics of the University of Groningen for its hospitality and
financial support and M.M.~Fern\'andez for her continuous support.
This work has been partially supported by the Spanish grant
FPA2000-1584.

\appendix
\section{The Reduction of Massive 11-Dimensional SUGRA to $d=5$}
\label{sec-reductiontod=5}

\subsection{Direct dimensional reduction of $D=11$ Supergravity on $T^{6}$}

\noindent
 The KK Ansatz for the Elfbein is

\begin{equation}
\left( \hat{e}_{\hat{\mu}}{}^{\hat{a}} \right) = 
\left(
\begin{array}{cr}
e^{\frac{1}{3}\varphi} e_{\mu}{}^{a} & 
e^{-\frac{1}{6}\varphi}L_{m}{}^{i}A^{m}{}_{\mu} \\
&\\
0             & e^{-\frac{1}{6}\varphi} L_{m}{}^{i}  \\
\end{array}
\right)
\, , 
\hspace{1cm}
\left(\hat{e}_{\hat{a}}{}^{\hat{\mu}} \right) =
\left(
\begin{array}{cr}
e^{-\frac{1}{3}\varphi}e_{a}{}^{\mu} & 
-e^{-\frac{1}{3}\varphi}A^{m}{}_{a}   \\
& \\
0             &  e^{\frac{1}{6}\varphi} L_{i}{}^{m}  \\
\end{array}
\right)\, , 
\end{equation}

\noindent
 and for the 3-form potential

\begin{equation}
  \begin{array}{rclrcl}
\hat{C}_{abc} & = & e^{-\varphi} C_{abc}\, ,&
\hat{C}_{abi} & = & e^{-\varphi/2} L_{i}{}^{m} 
B_{m\, ab}\, ,\\
& & & & & \\
\hat{C}_{aij} & = & 
L_{i}{}^{m} L_{j}{}^{n}V_{mn\, a}\, ,
\hspace{1.5cm}&
\hat{C}_{ijk} & = & e^{\varphi/2} L_{i}{}^{m} L_{j}{}^{n} L_{k}{}^{p} \partial_{a}\, a_{mnp}\, ,\\
  \end{array}
\label{eq:d-5-potential-flat}
\end{equation}

\noindent
 which, in curved components, are given in (\ref{eq:d8-C-curved}).

The 11-dimensional 4-form field strength decomposes as

\begin{equation}
\begin{array}{rclrcl}
\hat{G}_{abcd} & = & e^{-\frac{4}{3}\varphi} G_{abcd}\, ,&
\hat{G}_{abc\, i} & = & 
e^{-\frac{5}{6}\varphi} L_{i}{}^{m}\, H_{m\, abc}\, , \\
& & & & & \\
\hat{G}_{ab\, ij} & = & 
e^{-\frac{1}{3}\varphi}\, L_{i}{}^{m}L_{j}{}^{n}
\left[F_{mn\, ab} + a \epsilon_{mnp}\, F^{p}{}_{ab} \right]\, , 
\hspace{1cm} &
\hat{G}_{a\, ijk} & = & 
e^{\frac{1}{6}\varphi}\,  L_{i}{}^{m} L_{j}{}^{n} L_{k}{}^{p} \partial_{a}\, a_{mnp}\, ,\\
\end{array}
\label{eq:d-5-field-strengths-flat}
\end{equation}

\noindent
 such that the field strengths are

\begin{equation}
\begin{array}{rcl}
&&
  G_{\mu\nu\rho\sigma} = 4\partial_{[\mu}C_{\nu\rho\sigma]}
  +6B_{m\, [\mu\nu}F^{m}{}_{\rho\sigma]}\, ,\\
 & & \\
&&
  H_{m\, \mu\nu\rho} = 3\partial_{[\mu|}B_{m|\, \nu\rho]}
  +3V_{mn\, [\mu}F^{n}{}_{\nu\rho]}\, ,\\
 & & \\
&&
  F_{mn\, \mu\nu} = 2\partial_{[\mu|}V_{mn|\, \nu]}\, , \\
 & & \\
&&
  F^{m}{}_{\mu\nu} = 2\partial_{[\mu}A^{m}{}_{\nu]}\, .\\
\end{array}
\end{equation}

The action for the massless/ungauged 5-dimensional action is

\begin{equation}
\label{eq:d5action-1}
\begin{array}{rcl}
 S & = & 
{\displaystyle\int} d^{5}x\, \sqrt{|g|}\
\left\{
R +\frac{1}{2}\left(\partial\varphi \right)^{2}
+{\textstyle\frac{1}{4}} {\rm Tr} 
\left(\partial {\cal M} {\cal M}^{-1}\right)^{2}\, \right. \\
 & & \\
&&
\left.
-{\textstyle\frac{1}{4}} e^{-\varphi} 
F^{m} (A) {\cal M}_{mn} F^{n} (A)
-\frac{1}{2\cdot 4!} e^{-2\varphi} G^2
+\frac{1}{2\cdot 3!} e^{-\varphi} H_{m}{\cal M}^{mn} H_{n}\,
\right. \\
 & & \\
&&
\left.
  - \frac{1}{8}
    {\cal M}^{mp}
    {\cal F}_{mn}
    {\cal M}^{nq}
    {\cal F}_{pq}\,
+\frac{1}{18} e^{\varphi} {\cal M}^{mq} {\cal M}^{nr} {\cal M}^{ps}
     \partial a_{mnp}\, \partial a_{qrs}\, \right. \\
 & & \\
&&
 \left.
  -\frac{1}{2^{6} \cdot 3^{4}}
   \frac{\epsilon}{\sqrt{|g_{E}|}}
    \epsilon^{mnpqrs}
  \left[\,
     2\, G\, \partial a_{mnp}\, a_{qrs}
   + 12\, H_{m} {\cal F}_{np} a_{qrs}\, \right. \right. \\
 & & \\
&&
 \left.
  \left.
   +\, 24\, H_{m} \partial a_{npq} V_{rs}
   + 27\, {\cal F}_{mn} {\cal F}_{pq} V_{rs}
   + 36\, {\cal F}_{mn} \partial a_{pqr} B_{s}\, \right. \right. \\
 & & \\
&&
 \left.
  \left.
   +\, 4\, \partial a_{mnp}\, \partial a_{qrs} C\,
 \right]\,
\right\}\, .
\end{array}
\end{equation}

\noindent
 where

\begin{equation}
 {\cal F}_{mn} = F_{mn}(V) + a_{mnp} F^{p}(A)\, .
\end{equation}

In $d$ dimensions the Hodge-dual of a $p$-form potential is
a $d-p-2$ potential. We are interested in dualizing the three-form
and two-form fields $C$ and $B_{m}$ into scalar and vector potentials.
We get

\begin{equation}
\label{eq:G-dual}
\begin{array}{rcl}
 G &=& e^{2\varphi}\, {}^{*}\tilde{G}\, ,\\
& & \\
 H_{m} &=& e^{\varphi}\, {}^{*}\tilde{H}_{m}\, ,\\
\end{array}
\end{equation}

\noindent
where

\begin{equation}
\begin{array}{rcl}
 \tilde{G} &=& \partial a
            - \frac{1}{6^3}\, \epsilon^{mnpqrs}\,
              \partial a_{mnp}\, a_{qrs}\, ,\\
& & \\
 \tilde{H}_{m} &=& {\cal M}_{mn}\,
                 \left[
                   2\partial N^{n} - \frac{1}{36}\epsilon^{npqrsu}
                   {\cal F}_{pq} a_{rsu}
                   + a\, F^{n}(A)
                 \right]\, .
\end{array}
\end{equation}

Then, the action (\ref{eq:d5action-1}) in terms of the dual fields reads

\begin{equation}
\label{eq:d5action-2}
\begin{array}{rcl}
 S & = & 
{\displaystyle\int} d^{5}x\, \sqrt{|g|}\
\left\{
R 
+{\textstyle\frac{1}{4}} {\rm Tr} 
\left(\partial {\cal M} {\cal M}^{-1}\right)^{2}\,
+\frac{1}{2} \left( \partial\varphi \right)^{2}
+\frac{1}{2} e^{2\varphi} \tilde{G}^{2}\, \right. \\
 & & \\
&&
\left.
-{\textstyle\frac{1}{4}} e^{-\varphi} 
F^{m} (A) {\cal M}_{mn} F^{n} (A)
-\frac{1}{4} e^{\varphi} \tilde{H}_{m}{\cal M}^{mn} \tilde{H}_{n}\,
\right. \\
 & & \\
&&
\left.
  - \frac{1}{8}
    {\cal M}^{mp}
    {\cal F}_{mn}
    {\cal M}^{nq}
    {\cal F}_{pq}\,
+\frac{1}{18} e^{\varphi} {\cal M}^{mq} {\cal M}^{nr} {\cal M}^{ps}
     \partial a_{mnp}\, \partial a_{qrs}\, \right. \\
 & & \\
&&
 \left.
  -\frac{1}{2^{6} \cdot 3^{4}}
   \frac{\epsilon}{\sqrt{|g_{E}|}}
   \left[ \epsilon^{mnpqrs}
  \left(\,
    27 {\cal F}_{mn} {\cal F}_{pq} V_{rs}
  - 12 a_{mnp} F_{qr}(V)F^{u}(A)V_{us} \right)
   \right. \right. \\
 & & \\
&&
 \left.
  \left.
   - 6^4\, F^{m}(A) F_{mn}(V) N^{n}\,  \right]\,
\right\}\, .
\end{array}
\end{equation}









\subsection{Dimensional Reduction of Massive 11-Dimensional Supergravity}
\label{sec-massivesugra}

The decomposition of the 11-dimensional 3-form potential and 4-form field strength
are given in (\ref{eq:d-5-potential-flat}).
The 11-dimensional field strength decomposes as in (\ref{eq:d-5-field-strengths-flat}),
but now

\begin{equation}
\hat{G}_{a\, ijk} = 
e^{\frac{1}{6}\varphi}\,  L_{i}{}^{m} L_{j}{}^{n} L_{k}{}^{p} {\cal D}_{a}\, a_{mnp}\, ,\\
\end{equation}

\noindent
 this is, we have replaced $\partial$ by the $SO(6-l,l)$ covariant derivative ${\cal D}$.
There is also a new contribution from the eleven-dimensional field-strength, namely

\begin{equation}
\hat{G}_{ijkl} = 
e^{\frac{1}{6}\varphi}\,  L_{i}{}^{m} L_{j}{}^{n} L_{k}{}^{p} L_{l}{}^{q}
                          \left[
                            -3 \mathsf{Q}^{rs} a_{r[mn} a_{pq]s}
                          \right]\, ,\\
\end{equation}

\noindent
 which will contribute to the scalar potential.
The five-dimensional field strengths are now massive,
and are defined as in (\ref{eq:d-8-massive-field-strengths}).
The expressions for the massive gauge transformations are the same as those
in (\ref{eq:d-8-massive-gauge-transformations-1}) and
(\ref{eq:d-8-massive-gauge-transformations-2}).

Then, the $d=5$ massive action reads

\begin{equation}
\label{eq:d5massiveaction}
  \begin{array}{rcl}
S & = & {\displaystyle\int} d^{5}x \sqrt{|g_{E}|}\,
\left\{ 
R_{E} +\frac{1}{2}\left(\partial\varphi \right)^{2}
+{\textstyle\frac{1}{4}} {\rm Tr} 
\left({\cal D} {\cal M} {\cal M}^{-1}\right)^{2}\, \right. \\
 & & \\
&&
\left.
-{\textstyle\frac{1}{4}} e^{-\varphi} 
F^{m} (A) {\cal M}_{mn} F^{n} (A)
-\frac{1}{2\cdot 4!} e^{-2\varphi} G^2
+\frac{1}{2\cdot 3!} e^{-\varphi} H_{m}{\cal M}^{mn} H_{n}\,
\right. \\
 & & \\
&&
\left.
  - \frac{1}{8}
    {\cal M}^{mp}
    {\cal F}_{mn}
    {\cal M}^{nq}
    {\cal F}_{pq}\,
+\frac{1}{18} e^{\varphi} {\cal M}^{mq} {\cal M}^{nr} {\cal M}^{ps}
     {\cal D} a_{mnp}\, {\cal D} a_{qrs}\, - {\cal V}\, \right. \\
 & & \\
&&
 \left.
  -\frac{1}{2^{6} \cdot 3^{4}}
   \frac{\epsilon}{\sqrt{|g_{E}|}}
    \epsilon^{mnpqrs}\, 
  \left\{\,
     2\, G\, {\cal D} a_{mnp}\, a_{qrs}
   + 12\, H_{m} {\cal F}_{np} a_{qrs}\,
   + 24\, H_{m} {\cal D} a_{npq} V_{rs} \right. \right. \\
 & & \\
&&
 \left.
  \left.
   +\, 27\, {\cal F}_{mn} {\cal F}_{pq} V_{rs}
   + 36\, {\cal F}_{mn} {\cal D} a_{pqr} B_{s}
   + 4\, {\cal D} a_{mnp}\, {\cal D} a_{qrs} C\, \right. \right. \\
 & & \\
&&
 \left.
  \left.
  +\, 9 {\mathsf Q}^{vw}
  \left[
     2 \left( G V_{mn} + 4 H_{m} B_{n}
             + 2 {\cal F}_{mn} C \right) a_{pqv} a_{rsw}\, \right. \right. \right. \\
 & & \\
&&
 \left. \left. \left.
   +\, 2\left( G a_{mnp} + 12 H_{m} V_{np} + 18 {\cal F}_{mn} B_{p}
               + 3 {\cal D} a_{mnp} C \right) V_{qv} a_{rsw}\, \right. \right. \right. \\
 & & \\
&&
 \left. \left. \left.
   +\, \left( 4 H_{m} a_{npq} + 18 {\cal F}_{mn} V_{pq}
            + 9 {\cal D} a_{mnp} B_{q} \right) B_{v} a_{rsw}\, \right. \right. \right. \\
 & & \\
&&
 \left. \left. \left.
   +\, 4 \left( 2 H_{m} a_{npq} + 9 {\cal F}_{mn} V_{pq}
                + 6 {\cal D} a_{mnp} B_{q} \right) V_{rv} V_{sw}\, \right. \right. \right. \\
 & & \\
&&
 \left. \left. \left.
   +\, 12 \left( {\cal F}_{mn} a_{pqr}
              + 2 {\cal D} a_{mnp} V_{qr} \right) B_{v} V_{sw}\,
   + 3 {\cal D} a_{mnp} a_{qrs} B_{v} B_{w}\, \right] \right. \right. \\
 & & \\
&&
 \left. \left.
   +\, \frac{9}{10}\, \mathsf{Q}^{vw}\, \mathsf{Q}^{xy}
   \left[
    9 \left( 4 a_{mnp} a_{qrv} V_{sw}
            + 3 V_{mn} a_{pqv} a_{rsw} \right) B_{x} B_{y}\, \right. \right. \right. \\
 & & \\
&&
 \left. \left. \left.
   +\, 8 \left( 9 V_{mn} a_{pqv} V_{rw}
               + 16 B_{m} a_{npv} a_{qrw} \right) V_{sx} B_{y}\, \right. \right. \right. \\
 & & \\
&&
 \left. \left. \left.
   +\, 24 \left( 3 V_{mn} V_{pv} V_{qw} + 6 B_{m} a_{npv} V_{qw}
                + 12 C a_{mnv} a_{pqw} \right) V_{rx} V_{sy}\, \right. \right. \right. \\
 & & \\
&&
\left. \left. \left.
   +\, 36 C a_{mnv} a_{pqw} a_{rsx} B_{y}\,
  \right]\,
 \right\}
\right\}\, ,\\
\end{array}
\end{equation}

\noindent
where

\begin{equation}
 {\cal F}_{mn} = F_{mn}(V) + a_{mnp} F^{p}(A)\, 
\end{equation}

\noindent
with $F_{mn}(V)$ and $F^{m}(A)$ are the massive ones.
${\cal V}$ is the scalar potential, given by

\begin{equation}
\begin{array}{rcl}
{\cal V} &=& -\, \frac{1}{2}\,
          \left\{
             \left[ {\rm Tr}{({\cal M}\mathsf{Q})} \right]^2
            - 2\, {\rm Tr}{({\cal M}\mathsf{Q}{\cal M}\mathsf{Q})}\, \right.\\
& & \\
&&
\left.
             -\frac{1}{2}e^{\varphi}
              \left[
                (\mathsf{Q}{\cal M}\mathsf{Q})^{mq} {\cal M}^{nr} {\cal M}^{ps}
                - 2\, \mathsf{Q}^{mq} {\cal M}^{nr} \mathsf{Q}^{ps}
              \right] a_{mnp}a_{qrs}\, \right.\\
& & \\
&&
\left.
              -\frac{1}{24} e^{2\varphi}
                {\cal M}^{mr}{\cal M}^{ns}{\cal M}^{pt}{\cal M}^{qu}
                \left( 3\, a_{v[mn}a_{p]qw} \mathsf{Q}^{vw} \right)\,
                \left( 3\, a_{x[rs}a_{t]uy} \mathsf{Q}^{xy} \right)\,
           \right\}\, .
\end{array}
\label{eq:5-potential-1}
\end{equation}

\section{Extremizing the Scalar Potential}
\label{ap:potential-extrema}

The above potential is a complicated function on many independent
scalar variables plus the parameters of the mass matrix $\mathsf{Q}$,
which makes extremely difficult a complete study of its extrema.  Only
some extrema are known in the $SO(6-l,l)$ cases
\cite{Warner:1983vz,Warner:1984du,Khavaev:1998fb}.

In the general case of the series of gauged/masive supergravities that
we can generate from $d=11$, it is even more difficult to find and
study all the extrema of the potential. However, all the potentials
contain the ``universal term'' Eq.~(\ref{eq:universalterm}) that only
depends on $\mathcal{M}$ and the mass matrix $\mathsf{Q}$ and we can
try to systematically study it.

The universal term is a function of the dilaton $\varphi$ and the
matrix of scalars ${\cal M}$

\begin{equation}
 {\cal V} = {\cal V}(\varphi,{\cal M})\, ,
\end{equation}

\noindent
whose origin (in the family of gauged/massive supergravities that we
are studying) is the internal metric $G_{mn}$, which we have
decomposed as

\begin{equation}
 G_{mn} = e^{-\frac{2}{n}\varphi}\, {\cal M}_{mn}\, ,
\label{eq:internal-metric}
\end{equation}

\noindent
$n$ being the dimension of the compact space, the $n$-torus.

In the reduced theory the $Sl(n,\mathbb{R})/SO(n)$ matrix of scalars
${\cal M}_{mn}$ is subject to the constraint ${\rm det}({\cal M})=1$,
and this has to be taken into account to find its equations of motion.
the simplest way to do it is to calculate the equations of motion for
the unconstrained variables $G_{mn}$, and not ${\cal M}_{mn}$ and
$\varphi$ separately, and then use the chain rule:

\begin{equation}
 \frac{\delta S}{\delta G_{mn}} = 0
 \Rightarrow
{\textstyle\frac{n}{2}} {\cal M}^{mn} \frac{\delta S}{\delta \varphi}
     - \frac{\delta S}{\delta {\cal M}_{mn}} =0\, .
\end{equation}

\noindent
The potential ${\cal V}$ has to be extremized w.r.t.~$G_{mn}$ as well:

\begin{equation}
 \frac{\partial {\cal V}}{\partial G_{mn}} = 0\, ,
\,\,\,\,
\Rightarrow
{\textstyle \frac{n}{2}} {\cal M}^{mn} 
\frac{\partial {\cal V}}{\partial \varphi}
     - \frac{\partial {\cal V}}{\partial {\cal M}_{mn}} =0\, .
\label{eq:pot-2}
\end{equation}

\noindent
Contracting the equation above with ${\cal M}_{mn}$ we find

\begin{equation}
 \frac{\partial {\cal V}}{\partial \varphi}
 = {\textstyle\frac{2}{n}} \frac{1}{{\rm Tr}({\cal M}^2)}\, {\cal M}_{mn}
   \frac{\partial {\cal V}}{\partial {\cal M}_{mn}}\, ,
\,\,\,\,
\Rightarrow
 {\cal M}^{mn}{\cal M}_{pq}
 \frac{\partial {\cal V}}{\partial {\cal M}_{pq}}
 - {\rm Tr}({\cal M}^2)
 \frac{\partial {\cal V}}{\partial {\cal M}_{mn}} = 0\, .
\label{eq:pot-4}
\end{equation}

Let us consider the simplest case $\mathsf{Q}= g\,
\mathbb{I}_{n\times n}$.  In this case, Eq.~(\ref{eq:pot-4}) is

\begin{equation}
 {\cal M}^{mn} \left\{
                ({\rm Tr}{\cal M})^2
                - 2 {\rm Tr}({\cal M}^2)
               \right\}
               - {\rm Tr}({\cal M}^2)
               \left\{ {\rm Tr}({\cal M}) \delta^{mn}
                 - 2 {\cal M}^{mn}
               \right\} = 0\, ,
\label{eq:pot-4-Q=I}
\end{equation}

\noindent
which is solved by

\begin{equation}
 {\cal M} = \pm \mathbb{I}_{n\times n}\, , 
\end{equation}

\noindent
although only the positive sign gives a proper solution, consistent
with the signature of the 11-dimensional spacetime.

These vacua will in general have a non-trivial $\varphi$, except in
$d=5$ in which the universal term of the potential does not depend on
it and the vacuum solution is $AdS_{5}$.


\begin{thebibliography}{60}

\bibitem{Maldacena:1997re}
J.~M.~Maldacena,
Adv.\ Theor.\ Math.\ Phys.\  {\bf 2} (1998) 231
[Int.\ J.\ Theor.\ Phys.\  {\bf 38} (1999) 1113]
[arXiv:hep-th/9711200].

\bibitem{Boonstra:1998mp}
H.~J.~Boonstra, K.~Skenderis and P.~K.~Townsend,
JHEP {\bf 9901} (1999) 003
[arXiv:hep-th/9807137].

\bibitem{Aharony:1999ti}
O.~Aharony, S.~S.~Gubser, J.~M.~Maldacena, H.~Ooguri and Y.~Oz,
Phys.\ Rept.\  {\bf 323} (2000) 183
[arXiv:hep-th/9905111].

\bibitem{Freund:1980xh}
P.~G.~Freund and M.~A.~Rubin,
Phys.\ Lett.\ B {\bf 97} (1980) 233.

\bibitem{Duff:1983gq}
M.~J.~Duff and C.~N.~Pope,
ICTP/82/83-7
{\it Lectures given at September School on Supergravity and Supersymmetry, Trieste, Italy, Sep 6-18, 1982}.

\bibitem{Pilch:xy} K.~Pilch, P.~van Nieuwenhuizen and P.~K.~Townsend,
Nucl.\ Phys.\ B {\bf 242} (1984) 377.

\bibitem{Nastase:1999cb}
H.~Nastase, D.~Vaman and P.~van Nieuwenhuizen,
Phys.\ Lett.\ B {\bf 469} (1999) 96
[arXiv:hep-th/9905075].

\bibitem{Schwarz:qr}
J.~H.~Schwarz,
Nucl.\ Phys.\ B {\bf 226} (1983) 269.

\bibitem{Gunaydin:1984fk}
M.~G\"unaydin and N.~Marcus,
Class.\ Quant.\ Grav.\  {\bf 2} (1985) L11.

\bibitem{Kim:1985ez}
H.J.~Kim, L.J.~Romans and P.~van~Nieuwenhuizen, 
Phys.\ Rev.\ D\ {\bf 32} (1985) 389. 

\bibitem{Scherk:1979zr}
J.~Scherk and J.~H.~Schwarz,
Nucl.\ Phys.\ B {\bf 153} (1979) 61.

\bibitem{Salam:1984ft}
A.~Salam and E.~Sezgin,
Nucl.\ Phys.\ B {\bf 258} (1985) 284.

\bibitem{Chamseddine:1997nm}
A.~H.~Chamseddine and M.~S.~Volkov,
Phys.\ Rev.\ Lett.\  {\bf 79} (1997) 3343
[arXiv:hep-th/9707176].

\bibitem{Chamseddine:1997mc}
A.~H.~Chamseddine and M.~S.~Volkov,
Phys.\ Rev.\ D {\bf 57} (1998) 6242
[arXiv:hep-th/9711181].

\bibitem{Chamseddine:1999uy}
A.~H.~Chamseddine and W.~A.~Sabra,
Phys.\ Lett.\ B {\bf 476} (2000) 415
[arXiv:hep-th/9911180].

\bibitem{Bergshoeff:1996ui}
E.~Bergshoeff, M.~de Roo, M.~B.~Green, G.~Papadopoulos and P.~K.~Townsend,
Nucl.\ Phys.\ B {\bf 470} (1996) 113
[arXiv:hep-th/9601150].

\bibitem{Meessen:1998qm}
P.~Meessen and T.~Ort\'{\i}n,
Nucl.\ Phys.\ B {\bf 541} (1999) 195
[arXiv:hep-th/9806120].

\bibitem{Gheerardyn:2001jj}
J.~Gheerardyn and P.~Meessen,
Phys.\ Lett.\ B {\bf 525} (2002) 322
[arXiv:hep-th/0111130].

\bibitem{Cowdall:2000sq}
P.~M.~Cowdall,
arXiv:hep-th/0009016.

\bibitem{Hull:2002wg}
C.~M.~Hull,
arXiv:hep-th/0203146.

\bibitem{Bergshoeff:2002mb}
E.~Bergshoeff, U.~Gran and D.~Roest,
Class.\ Quant.\ Grav.\  {\bf 19} (2002) 4207
[arXiv:hep-th/0203202].

\bibitem{Nishino:2002zi}
H.~Nishino and S.~Rajpoot,
arXiv:hep-th/0207246.

\bibitem{Bergshoeff:2002nv}
E.~Bergshoeff, T.~de Wit, U.~Gran, R.~Linares and D.~Roest,
arXiv:hep-th/0209205.

\bibitem{Louis:2002ny}
J.~Louis and A.~Micu,
Nucl.\ Phys.\ B {\bf 635} (2002) 395
[arXiv:hep-th/0202168].

\bibitem{deWit:1981eq}
B.~de Wit and H.~Nicolai,
Phys.\ Lett.\ B {\bf 108} (1982) 285.

\bibitem{deWit:1982ig}
B.~de Wit and H.~Nicolai,
Nucl.\ Phys.\ B {\bf 208} (1982) 323.

\bibitem{Pernici:xx}
M.~Pernici, K.~Pilch and P.~van Nieuwenhuizen,
Phys.\ Lett.\ B {\bf 143} (1984) 103.

\bibitem{Gunaydin:1984qu}
M.~G\"unaydin, L.~J.~Romans and N.~P.~Warner,
Phys.\ Lett.\ B {\bf 154} (1985) 268.

\bibitem{Pernici:ju}
M.~Pernici, K.~Pilch and P.~van Nieuwenhuizen,
Nucl.\ Phys.\ B {\bf 259} (1985) 460.

\bibitem{Freedman:ra}
D.~Z.~Freedman and J.~H.~Schwarz,
Nucl.\ Phys.\ B {\bf 137} (1978) 333.

\bibitem{Townsend:1983kk}
P.~K.~Townsend and P.~van Nieuwenhuizen,
Phys.\ Lett.\ B {\bf 125} (1983) 41.

\bibitem{Romans:tz}
L.~J.~Romans,
Phys.\ Lett.\ B {\bf 169} (1986) 374.

\bibitem{Townsend:xs}
P.~K.~Townsend, K.~Pilch and P.~van Nieuwenhuizen,
Phys.\ Lett.\  {\bf 136B} (1984) 38
[Addendum-ibid.\  {\bf 137B} (1984) 443].

\bibitem{Townsend:ad}
P.~K.~Townsend,
in {\it Quantum Field Theory and Quantum Statistics, Vol. 2, 299-308}
Batalin, I.A. et al (Ed.).

\bibitem{Bautier:1997yp}
K.~Bautier, S.~Deser, M.~Henneaux and D.~Seminara,
Phys.\ Lett.\ B {\bf 406} (1997) 49
[arXiv:hep-th/9704131].

\bibitem{Deser:1997gm}
S.~Deser,
arXiv:hep-th/9712064.

\bibitem{Deser:1998xi}
S.~Deser,
arXiv:hep-th/9805205.

\bibitem{Pernici:zw}
M.~Pernici, K.~Pilch, P.~van Nieuwenhuizen and N.~P.~Warner,
Nucl.\ Phys.\ B {\bf 249} (1985) 381.

\bibitem{Hull:2002cv}
C.~M.~Hull,
arXiv:hep-th/0204156.

\bibitem{Bergshoeff:1997ak}
E.~Bergshoeff, Y.~Lozano and T.~Ort\'{\i}n,
Nucl.\ Phys.\ B {\bf 518} (1998) 363
[arXiv:hep-th/9712115].

\bibitem{Alonso-Alberca:2000gh}
N.~Alonso-Alberca, P.~Meessen and T.~Ort\'{\i}n,
Nucl.\ Phys.\ B {\bf 602} (2001) 329
[arXiv:hep-th/0012032].

\bibitem{Hull:1994ys}
C.~M.~Hull and P.~K.~Townsend,
Nucl.\ Phys.\ B {\bf 438} (1995) 109
[arXiv:hep-th/9410167].

\bibitem{Gunaydin:1986cu}
M.~Gunaydin, L.J.~Romans and N.P.~Warner,
Nucl.\ Phys.\ B {\bf 272} (1986) 598.

\bibitem{Cvetic:2000nc}
M.~Cvetic, H.~Lu, C.N.~Pope, A.~Sadrzadeh and T.A.~Tran,
Nucl.\ Phys.\ B {\bf 586} (2000) 275
[arXiv:hep-th/0003103].

\bibitem{Cremmer:1978km}
E.~Cremmer, B.~Julia and J.~Scherk,
Phys.\ Lett.\ B {\bf 76} (1978) 409.

\bibitem{Haack:2001jz}
M.~Haack and J.~Louis,
Phys.\ Lett.\ B {\bf 507} (2001) 296
[arXiv:hep-th/0103068].

\bibitem{Warner:1983vz}
N.P.~Warner,
Phys.\ Lett.\ B {\bf 128} (1983) 169.

\bibitem{Warner:1984du}
N.P.~Warner,
Nucl.\ Phys.\ B {\bf 231} (1984) 250.

\bibitem{Khavaev:1998fb}
A.~Khavaev, K.~Pilch, and N.~Warner,
Phys.\ Lett.\ B {\bf 487} (2000) 14
[arXiv:hep-th/9812035].

\bibitem{kn:KSMCH} D.~Kramer, H.~Stephani, M.~MacCallum and E.~Herlt,
                   {\sl Exact Solutions of Einstein's Field Equations},
                   Cambridge University Press, Cambridge, U.K. (1980).



\end{thebibliography}
\end{document}